\documentclass[english, letterpaper, 11pt]{article}

\usepackage{fouriernc}
\usepackage[T1]{fontenc}

\usepackage[margin=1in]{geometry}
\usepackage{amssymb}
\usepackage{amsmath} % For theorems
\usepackage{mathrsfs}
\usepackage{bbm} % For double-stroke unity
\usepackage{scrextend} % For \addmargin{}
\usepackage{natbib} 
\usepackage{graphicx} % Allows inclusion of figures
\usepackage{color}
\usepackage{booktabs} % Allows the use of \toprule, \midrule and \bottomrule in tables
\usepackage[labelsep=period]{caption} % Allows `\\' in captions
\usepackage[normalem]{ulem} % Allows crossing out text
\usepackage[hang,flushmargin]{footmisc}
\usepackage{rotating} % Allows landscape-oriented tables
\usepackage{scalefnt} % Allows scaling fontsize
\usepackage{setspace} % Allows doublespacing
\usepackage{cite}
\usepackage{threeparttable}
\usepackage{caption}
\usepackage{multirow}
\usepackage{titlesec}

\titleformat{\section}
{\normalfont\Large\scshape}{\thesection.}{1em}{}
\titleformat{\subsection}
{\normalfont\large\itshape}{\thesubsection.}{1em}{}
\titleformat{\subsubsection}
{\normalfont\normalsize\itshape}{\thesubsubsection.}{1em}{}

\numberwithin{equation}{section}

\allowdisplaybreaks

\usepackage{authblk} % Author affiliation package

\numberwithin{equation}{section}
\newsavebox{\overlongequation}

\begin{document}

\title{\Huge Detecting Learning \textit{by} Exporting \\ and \textit{from} Exporters%: \\ Plant-Level Evidence from Chilean Manufacturing
\thanks{\textit{Acknowledgments}: We would like to thank Subal Kumbhakar and seminar participants at UNLV for many insightful comments on an earlier draft of the paper. \ \textit{Correspondence}: Emir Malikov, Lee Business School, University of Nevada, Las Vegas, Las Vegas, NV 89154-6005. Email: emir.malikov@unlv.edu.}}
\author[1]{\sc \vspace{0.2cm} Jingfang Zhang}
\author[2]{\sc  Emir Malikov}
\affil[1]{\small University of Kentucky}
\affil[2]{\small University of Nevada, Las Vegas}

\date{\small February 15, 2023}
\maketitle

%\vspace{\baselineskip}

\begin{abstract} \small
\noindent Existing literature at the nexus of firm productivity and export behavior mostly focuses on ``learning by exporting,'' whereby firms can improve their performance by engaging in exports. Whereas, the secondary channel of learning via cross-firm spillovers from exporting peers, or ``learning from exporters,'' has largely been neglected. Omitting this important mechanism, which can benefit both exporters and non-exporters, may provide an incomplete assessment of the total productivity benefits of exporting. In this paper, we develop a unified empirical framework for productivity measurement that explicitly accommodates both channels. To do this, we formalize the evolution of firm productivity as an export-controlled process, allowing future productivity to be affected by both the firm's own export behavior as well as export behavior of spatially proximate, same-industry peers. This facilitates a simultaneous, ``internally consistent'' identification of firm productivity and the corresponding effects of exporting. We apply our methodology to a panel of manufacturing plants in Chile in 1995--2007 and find significant evidence in support of both direct and spillover effects of exporting that substantially boost the productivity of domestic firms.

\vspace{\baselineskip}



\end{abstract}

\onehalfspacing
\thispagestyle{empty} \addtocounter{page}{0}
\clearpage

% ------------------------------------------------------------------------------------------
% ------------------------------------------------------------------------------------------

\section{Introduction}\label{sec:introduction}

Governments in both developing and developed countries commonly pursue policies aimed at promoting exports. In addition to boosting aggregate demand, such policies are also routinely justified by arguing that domestic exporters benefit from export-driven productivity improvements via absorption of new technologies from abroad, learning of international best practices that lead to improved business processes, productivity enhancements driven by exposure to more competition, scale effects, quality and variety effects, etc. This productivity-enhancing mechanism is usually referred to as ``learning by exporting'' (LBE) \citep[e.g.,][]{clerides1998learning, aw2000productivity, delgado2002firm,baldwin2004trade, van2005exporting,de2007exports,de2013detecting}. Such export-related productivity gains are facilitated by the firm's \textit{own} direct access to foreign customers, partners and rivals. 

Domestic firms, however, can learn not only from their own export experiences but also from their local \textit{peers} who engage in exports, and this indirect learning opportunity is available to both exporters and non-exporters. These external export-driven productivity spillovers are a type of cross-firm peer effects, which we refer to as ``learning from exporters'' (LFE), and effectively capture the secondary productivity effects of export engagement. Such cross-firm spillovers may occur via labor turnover, learning by imitation, exposure to affiliate and/or competitor products, etc.~\citep[see][]{greenaway2004domestic, sala2015export}. For instance, the movement of workers from exporting firms to other domestic firms may facilitate the dispersion of tacit knowledge about more innovative/efficient foreign technologies or institutional knowledge about foreign markets, which may help these firms improve their productivity too. 

Taking the indirect cross-firm productivity effect of export engagement for granted, as customarily done in literature on the productivity--exports nexus, will likely underestimate the total productivity benefits of exporting. Besides, omitting this important mechanism may also contaminate the measurement of more traditional LBE effects on firm productivity because it may lead to an endogeneity-inducing omitted variable bias. In this paper, we contribute to the literature by extending \citet{de2013detecting} to develop a unified empirical framework for productivity measurement that explicitly accommodates both the direct LBE channel taking place \textit{within} the firm as well as the indirect LFE channel working \textit{between} spatially and industrially proximate firms, which we then apply to a panel of manufacturing plants in Chile in 1995--2007.

While the literature on (within-firm) LBE effects is rather well-established  \citep[e.g.,][]{kunst1989exports,aw1995productivity, bernard1999exceptional,baldwin2003export,greenaway2004exporting,keller2004international, 
blalock2004learning,de2007exports,de2013detecting,wagner2007exports,Salomon2008, park2010exporting, aw2011r, kasahara2013productivity,Manjon2013}, the empirical analysis of external effects of exporting on \textit{productivity} in the industry beyond exporter firms is practically non-existent. Existing empirical work on ``export spillovers'' focuses mainly on how average export participation in the industry affects the export status or marginal cost of non-exporters nearby \citep[e.g.,][]{aitken1997spillovers, clerides1998learning,bernard2004some, greenaway2004domestic,  greenaway2008exporting,koenig2009agglomeration,koenig2010local,alvarez2013previous,poncet2013french} whereas the productivity implications of export spillovers in the domestic industry remain understudied, essentially being limited to three studies \citep[see][]{wei2006productivity,alvarez2008exporting,jung2010}.\footnote{\citet{wei2006productivity} examine productivity spillovers from exports in China's manufacturing, \citet{alvarez2008exporting} use older Chilean manufacturing data, and \citeauthor{jung2010}'s (2010)  dissertation studies Korean manufacturing plants. All three, however, employ empirical strategies that are affected by the econometric issues we discuss here.} 

%Prior attempts to examine them are essentially limited to just two studies: \citet{wei2006productivity} who examine productivity spillovers from exports in China's manufacturing and \citet{alvarez2008exporting} who use older Chilean manufacturing data. 
%\footnote{Both studies, however, employ empirical strategies that are not only overly restrictive or ``internally inconsistent'' in their modeling of firm productivity (or both) but are also seriously hindered by well-known identification problems associated with the production function estimation.}

These empirical analyses of productivity spillovers from exporting are typically done in one of two ways.\footnote{In fact, this is also the case for studies of spillovers from imports, R\&D, FDI and other production-related firm activities in the market; see \citet{malikovzhao2019} for more discussion.} The first approach is two-step, whereby one first estimates unobserved firm productivity via standard proxy methods while ignoring the dependence of the former on exports under the assumption of \textit{exogenous} first-order Markov evolution of productivity and then examines spillover effects in a second step by (linearly) regressing the already estimated firm productivity on the average export spillover exposure \citep[e.g.,][]{alvarez2008exporting,jung2010}. Taken at its face value, such a second-step analysis is problematic because it contradictorily postulates the existence of an endogenous exporting-productivity relationship that is at odds with the assumption about firm productivity being purely autoregressive in the first stage. \citet{de2013detecting} makes the same argument in his critique of two-step analyses of LBE, but it is equally important in the context of spillovers via LFE. Aside from the econometric complications arising from its inherently contradictory setup, the two-step approach also cannot provide structurally meaningful interpretation of LFE effects on productivity because of its inability to distinguish between different data-generating processes that all\textemdash and not only the one with non-zero LFE\textemdash can produce a positive relation between the average export participation of industry and firm productivity.

We avoid this internal inconsistency in our model by explicitly accounting for potential export spillover-induced productivity improvements during the estimation of firm productivity, which is partly what enables us to consistently estimate LFE effects along with other production-function components. Alternatively, estimates of both productivity and export effects thereon can be biased due to omission of the relevant measurement of the peers' exporting activities from the productivity-proxy function that are correlated with the firm's own export behavior as well as quasi-fixed inputs (via its latent productivity). Empirical findings of external export spillovers based on a two-step estimation procedure may therefore be spurious.

The other popular approach to testing export spillovers involves a single step and is based on estimation of the \citet{griliches1979}-style ``augmented production function'' which, besides the conventional inputs, also admits\textemdash usually in a linear fashion\textemdash measures of the firm's own and others' exporting activities \citep[e.g.,][]{wei2006productivity}. Although such an approach explicitly recognizes the existence of both LBE and LFE effects on productivity during the estimation, it does so by restrictively assuming that productivity effects of exporting are homogeneous across all firms, e.g., no matter their productivity levels. Besides linearity of the log Cobb-Douglas form that rules out heterogeneous effects, such an approach is also problematic because it assumes that the relationship between exporting and productivity is deterministic and, more importantly, it implies the possibility for a unit-elastic substitution between inputs and export variables \citep[see][for more on these points]{de2013detecting}.

To improve upon the above methods, in this paper we seek to measure and test the direct LBE and indirect LFE effects on productivity in a consistent structural framework of firm production. Building upon \citet{doraszelski2013r} and \citet{de2013detecting}, we formalize the evolution of firm productivity as an endogenous export-controlled process, where we explicitly accommodate two potential channels \textemdash internal and external (direct and indirect)\textemdash by which exporting may impact productivity of domestic firms. Specifically, we allow the firm's future productivity to be affected not only by its own past export behavior but also by that of its spatially proximate peers in the industry. This facilitates a simultaneous, internally consistent identification of firm productivity and corresponding LBE and LFE effects.\footnote{While our approach is related to that by \citet{malikovzhao2019} whose recent work also concerns an internally consistent measurement of cross-firm productivity spillovers, it is distinct in that we focus on the identification of ``contextual'' spillover effects\textemdash in the \citet{manski1993} nomenclature\textemdash of peers' \textit{activities} (namely, exporting) on productivity whereas they consider the measurement of ``endogenous'' effects of peers' \textit{productivity}. The type of spillovers that we study here is more predominant in the literature, particularly in the context of exporting. }

Given the objective of our paper, in our analysis we abstract away from other plausible and no less important channels of cross-firm interactions that may lead to productivity spillovers, including local agglomeration, R\&D externalities via technology/knowledge sharing, spillovers along the supply chains and material-product connections, and focus exclusively on spillovers from exporting. Similar to \citeauthor{de2013detecting}'s (2013) approach, ours is also silent on the exact theoretical mechanism by which LBE and LFE occur. Instead of pinpointing specific channels which is predictably more demanding on data and experiment design, we pursue a simpler but more feasible goal of testing for the presence of these export-driven effects on productivity in general.

To achieve identification in the wake of endogeneity-inducing  correlation of input allocations and, potentially, exporting with unobserved firm productivity \citep{griliches1995production}, we rely on the popular proxy variable technique. Our identification strategy utilizes a structural link between the parametric production function and the firm's first-order condition for static inputs which helps us circumvent \citeauthor{ackerberg2015identification}'s (2015) and \citeauthor{gandhi2018identification}'s (2018) non-identification critiques of conventional proxy-based estimators \`a la \citet{olley1992dynamics} and  \citet{levinsohn2003estimating}. In addition, owing to a nonparametric treatment of the firm productivity process, our model enables us to accommodate heterogeneity in productivity effects of exporting across firms. This also lets us explore potential nonlinearities in LBE and LFE effects whereby they can interact with each other as well as, more importantly, with the firm's own productivity thus indirectly allowing for conditioning on the firm's capacity to learn given its proximity to the technology frontier. To this end, not only do we provide a more comprehensive picture of the productivity effects of exporting, but we do so in a robust way by dealing with internal inconsistency and identification problems prevalent in prior literature.

We study the productivity-enhancing effects of exporting using plant-level data on Chilean manufacturers during the 1995--2007 period, with exporters accounting for about 21\% of the sample. Using our semiparametric methodology, we find that exporters enjoy a statistically significant productivity premium over non-exporters along the entire distribution of productivity which, however, can be reflective of both learning and self-selection 	effects. Zooming in on the within-firm learning effects of exporting, we find significant evidence of both LBE and LFE. Overall, the LBE productivity effect is statistically significant for 93\% of firms in our main specification. On average, the size of LFE effect (cumulatively from all peers) is commensurate to that of LBE but, at the observation level, LFE is significantly non-zero for 69\% of plants only, thereby suggesting that the indirect productivity-boosting effect of exporting is less prevalent in Chilean manufacturing than the direct learning effect taking place within the firm. We also document that the LFE effect is stronger for the firms who also export themselves. This empirical evidence therefore suggests that exporters benefit from the exposure to peer exporters in their local industry more than non-exporters, plausibly because there may be complementarities between internal/direct and external/indirect learning from export experiences. Using simple back-of-envelope calculations, we find that long-run total LBE and LFE effects on mean firm productivity are, respectively, 0.70\% and 0.63\% per percentage point of export intensity. Thus, in the long-run equilibrium a permanent 10 percentage point increase in mean export intensity of \textit{all} firms in the local industry is roughly estimated to produce a sizable 13.3\% boost to average firm productivity through both direct LBE and indirect LFE channels.

The rest of the paper is organized as follows. Section \ref{sec:framework} presents the conceptual framework. Section \ref{sec:empiricalframework} describes our identification and estimation procedure. The data are discussed in Section \ref{sec:data}. We report the empirical results in Section \ref{sec:results}. Section \ref{sec:conclusion} concludes.

% ------------------------------------------------------------------------------------------
% ------------------------------------------------------------------------------------------

\section{Conceptual Framework}\label{sec:framework}

Consider the firm $i( = 1, \dots, n)$ at time $t(= 1,\dots,T)$. Following the convention of productivity literature \citep[e.g.,][]{olley1992dynamics, blundell2000gmm, levinsohn2003estimating,de2012markups, doraszelski2013r, ackerberg2015identification,konings2015impact,jin2019financial}, we assume
that the firm employs physical capital $K_{it}$, labor $L_{it}$ and an intermediate input such as materials $M_{it}$ to produce the output $Y_{it}$
via the Cobb-Douglas production technology subject to the Hicks-neutral productivity:
\begin{equation}\label{eq:prodfun}
Y_{it}=A_{0}K_{it}^{\alpha_{K}}L_{it}^{\alpha_{L}}M_{it}^{\alpha_{M}}\exp\left\{ \omega_{it}+\eta_{it}\right\} ,
\end{equation}
where $A_0$ is a scalar constant; $(\alpha_K,\alpha_L,\alpha_M)'$ are the input elasticities; $\omega_{it}$ is the firm's persistent productivity (TFP) which is known
to the firm at time $t$ but unknown to an econometrician; and $\eta_{it}$ is a
random \textit{i.i.d.}~productivity shock such that $ E[\eta_{it}|\mathcal{I}_{it}]=E[\eta_{it}]=0$, where $\mathcal{I}_{it}$ is the $i$th firm's information set in period $t$.%\footnote{We have also experimented with adding the time trend and its square term into the production function to control for temporal change. Our findings about the LBE and LFE productivity effects are largely unchanged and continue to hold. For simplicity sake, we have opted for a more parsimonious specification.}

As in many studies in productivity literature \citep[e.g.,][]{gandhi2018identification,tsionas2019bayesian, hou2020productivity,malikovzhao2019,malikov2023}, physical capital $K_{it}$ and labor $L_{it}$ are said to be subject to adjustment frictions, and the firm optimizes them dynamically  at time $t-1$ rendering these inputs predetermined quasi-fixed state variables. The materials input $M_{it}$ is freely varying and determined by the firm statically at time $t$. Both $K_{it}$ and $L_{it}$ follow their respective laws of motion:
\begin{equation}\label{eq:kllaw}
K_{it}=I_{it-1}+\left(1-\delta\right)K_{it-1}\quad \text{and}\quad L_{it}=H_{it}+L_{it-1},
\end{equation}
where $I_{it}$, $H_{it}$ and $\delta$ respectively denote the gross investment, net hiring  and capital depreciation rate of the firm $i$ in period $t$. We assume that the risk-neutral firm faces perfectly competitive output and input markets and seeks to maximize a discounted stream of the expected life-time profits subject to its state variables and expectations about the market structure variables including prices that are common to all firms.

In this paper, our principal interest is in the measurement of internal and external productivity effects of exporting in a domestic industry. Instead of modeling the firm's export behavior in a discrete fashion by focusing on its ``status'' as popularly done in the literature \citep[e.g.,][]{blalock2004learning,van2005exporting, amiti2007trade, kasahara2013productivity}, we formalize the firm's exporting in a richer, continuous framework along the lines of \citet{de2007exports} and  \citet{malikov2017Estimation}. We rely on the firm's export intensity as a measure of its own export behavior as well as to model its exposure to peer exporters in the industry. 
Concretely, let $X_{it}\in[0,1]$ denote the firm's export intensity defined as the nominal share of its total output produced for export abroad, with its boundary values corresponding to wholly domestic and fully export-oriented firms. 

We conceptualize the firm's exporting decisions as a choice of the degree of its export orientation $X_{it}$. Given the documented persistence of export experience over time, we formalize these decisions as dynamic but, unlike other dynamic production choices by the firm  in our model such as capital investment or hiring, they need not be subject to adjustment frictions that would render them delayed and, hence, predetermined.\footnote{To make these distinctions clearer: if the optimal decision concerning production in period $t$ is affected by its history, then that decision is said to be ``dynamic.'' If, due to adjustment frictions, a decision concerning production in period $t$ is effectively made at $t-1$, then we say it is ``predetermined.'' In this nomenclature, $K_{it}$ and $L_{it}$ are dynamic and predetermined, whereas $X_{it}$ is dynamic but chosen at time $t$.} This assumption is important because it does \textit{not} rule out a contemporaneous correlation between firm productivity $\omega_{it}$ and its export orientation $X_{it}$. Thus, the firm's export intensity $X_{it}$ evolves according to the following dynamic process:
\begin{equation}\label{eq:xlaw}
X_{it}=\mathcal{X}_{it}+ X_{it-1},
\end{equation}
where $\mathcal{X}_{it}$ regulates the endogenous adjustment in the degree of export orientation $X_{it}$ at $t$.
	
The firm's dynamic optimization problem is then described by the following Bellman equation:
\begin{align}\label{eq:bellman}
	\mathbb{V}_{t}\big(\Xi_{it}\big) = \max_{I_{it},H_{it},\mathcal{X}_{it}} \Big\{ &\Pi_{t}(\Xi_{it}) -
	\text{C}^{I}_{t}(I_{it},K_{it}) -
	\text{C}^{H}_{t}(H_{it},L_{it})-
	\text{C}^{X}_{t}(\mathcal{X}_{it},X_{it-1}) + \mathbb{E}\Big[\mathbb{V}_{t+1}\big(\Xi_{it+1}\big) \Big| \Xi_{it},I_{it},H_{it},\mathcal{X}_{it}\Big]\, \Big\} ,
\end{align}
where $\Xi_{it}=(\omega_{it},K_{it},L_{it},X_{it-1})'\in \mathcal{I}_{it}$ are the state variables; $\Pi_{t}(\Xi_{it})$ is the restricted profit function derived as a value function corresponding to the static optimization problem in \eqref{eq:profitmaximaztion}; and $\text{C}^{\kappa}_t(\cdot)$ is the cost of adjustments in capital ($\kappa=I$), labor ($\kappa=H$) and exporting ($\kappa=X$). From the laws of motion in \eqref{eq:kllaw}--\eqref{eq:xlaw}, in the above dynamic problem the firm's exporting behavior $X_{it+1}$ is chosen (via $\mathcal{X}_{it+1}$) in time period $t+1$ unlike the amounts of dynamic inputs $K_{it+1}$ and $L_{it+1}$ that are chosen by the firm in time period $t$ (via $I_{it}$ and $H_{it}$, respectively). Solving \eqref{eq:bellman} for $I_{it}$, $H_{it}$ and $\mathcal{X}_{it}$ yields their respective optimal policy functions.

Next we formalize the productivity effects of exports. We do so by extending \citeauthor{de2013detecting}'s (2013) framework to accommodate not only the more traditional direct LBE effects but also to allow for indirect effects via learning from exporting peers. That is, we explicitly model two potential channels\textemdash internal and external (direct and indirect)\textemdash by which exporting may impact the productivity of domestic firms. 
The first channel, referred to as ``learning by exporting,'' takes place \textit{within} the firm internally and is commonly attributed to the exporter firm's absorption of new technologies from abroad, learning of international best practices that lead to  improved manufacturing processes, productivity enhancements driven by the exposure to more competition, scale effects, quality and variety effects, etc. These export-related productivity gains are facilitated by the firm's \textit{own} direct access to foreign customers and rivals. 

The second channel is less obvious and oftentimes left unaccounted. Domestic firms (both the exporters and non-exporters) can learn not only from their own export behavior but also indirectly from their exporting \textit{peers}'. These external export-driven productivity spillovers are a type of \textit{cross-firm} peer effects, which we refer to as ``learning from exporters,'' and effectively capture the secondary productivity effects of export engagement. Such cross-firm spillovers may arise through pooling of workers, informal contacts (e.g., attendance of trade shows, exposure to affiliate and/or competitor products and marketing, learning by imitation, customer-supplier discussions) or more formal reverse engineering. For instance, by monitoring successful exporting peers' market behavior, both in domestic and foreign markets, domestic firms can imitate and then adopt their business strategies to boost own productivity. Alternatively, the movement of labor from exporting firms to other domestic firms may facilitate the dispersion of tacit knowledge about more innovative/efficient foreign technologies and better business practices or institutional knowledge about foreign markets, which may help hiring firms increase their productivity. 

To capture export-driven productivity spillovers, we proxy each firm's exposure to exporters in the industry using the average export intensity of its spatially proximate \textit{peers} operating in the same industry defined as
\begin{equation}
\overline{X}_{it}=\sum_{j\ne i} p_{ijt}X_{jt},
\end{equation}
where $\{p_{ijt};\ j(\ne i)=1\dots,n\}$ are the peer-firm weights identifying exporters in the firm $i$'s industry and spatial locality in period $t$. In our baseline specification, we construct the peer connection weights as $p_{ijt}={1}(j\in\mathcal{L}_{it})\big/\sum_{k\left(\neq i\right)=1}^{n}{1}(k\in\mathcal{L}_{it})$,
where $\mathcal{L}_{it}$ denotes a set of firms that
are in the same industry and geographical region as is the firm $i$ in time period $t$. Thus, for the $i$th firm at time $t$ this weighting scheme assigns a uniform weight of $1/n_{it}$ to all other firms from the same industry in the same region, where $n_{it}$ is the total number of such peer firms. This scheme is based on the conventional argument that geographical proximity and industry play a central role in productivity spillovers.  For example, it is technologically easier and less costly for firms to monitor and mimic strategies of other exporters that operate within the same industry and region. This is also in line with the literature on export spillovers \citep[e.g., see][]{bernard2004some, greenaway2008exporting, koenig2009agglomeration, koenig2010local, poncet2013french}. We weight all peers equally, given that their export intensity which is being averaged to compute $\overline{X}_{it}$ is already measured relative to the scale of  their production.

Note that our export exposure measure is firm-specific because it excludes the $i$th firm. Thus,
$\overline{X}_{it}$ captures
the \textit{external} export orientation of the local industry which firm $i$ is exposed to. This measure varies across both the firms and time. While closely related, $\overline{X}_{it}$ is therefore not the ``grand'' industry average but the peer average in the industry. This distinction is crucial for the separable identification of LBE and LFE effects (more on this later). 

We model firm productivity evolution as a controlled first-order Markov process, whereby we allow firm $i$ to improve its future productivity not only via learning by exporting but also via learning from exporting peers. Generalizing \citeauthor{doraszelski2013r}'s (2013)
and \citeauthor{de2013detecting}'s (2013) formulations to include the indirect cross-firm effects of exporting, the productivity process is
\begin{equation}\label{eq:productivitylaw}
\omega_{it}=h\left(\omega_{it-1},X_{it-1},\overline{X}_{it-1}\right)+\zeta_{it},
\end{equation}
where $h(\cdot)$ is the conditional mean function of $\omega_{it}$; and $\zeta_{it}$
is a zero-mean random innovation in persistent productivity that is unanticipated by the firm at $t-1$: 
$ E\left[\zeta_{it} | \mathcal{I}_{it-1}\right]=0$. The LBE and LFE effects can then be measured as $LBE_{it}={\partial E\left[ \omega_{it}|\cdot\right]}\big/{\partial X_{it-1}}$ and $LFE_{it}={\partial E\left[ \omega_{it}|\cdot\right]}\big/{\partial\overline{X}_{it-1}}$, respectively. 

Identification of the productivity effects of exporting in our model is based on several structural timing assumptions. The evolution process in (\ref{eq:productivitylaw}) assumes that both the internal and external learning associated with exports occurs with a delay which is why the dependence of $\omega_{it}$ on controls is lagged implying that export-driven improvements in firm productivity take a period to materialize. Such a timing assumption is quite common in the LBE literature \citep[e.g.,][]{van2005exporting,de2013detecting,malikov2017Estimation,malikov2021jom, malikov2023}. In $E[\zeta_{it} | \mathcal{I}_{it-1}]=0$, we also assume that the firm does not experience changes in exporting in light of expected \textit{future} innovations in its productivity. This rules out the firm's ability to systematically predict future productivity shocks. Instead, the productivity process in \eqref{eq:productivitylaw} says that firms anticipate the effect of their export experience $X_{it-1}$ on future productivity $\omega_{it}$ when adjusting the former in period $t-1$, and the conditional mean $E[\omega_{it} | \omega_{it-1}, X_{it-1},\overline{X}_{it-1}]$ captures that \textit{expected} productivity. But the \textit{actual} firm productivity at time $t$ also includes an unanticipated innovation $\zeta_{it}$. In essence, the error term $\zeta_{it}$ represents unpredictable uncertainty that is naturally associated with any productivity-modifying learning activities such as chance in discovery, success in implementation, etc. The  innovation $\zeta_{it}$ is realized after $X_{it-1}$ will have already been chosen. This structural timing assumption about the arrival of $\zeta_{it}$, which renders the firm's \textit{past} export experience mean-orthogonal to a random innovation at time $t$, helps us identify both the direct learning and external spillover effects. In Appendix \ref{sec:appx_selection}, we also discuss how inclusion of the lagged firm productivity in \eqref{eq:productivitylaw} indirectly allows us to, at least partly, control for self-selection of firms into exporting based on their productivity levels.

Since, in the productivity process \eqref{eq:productivitylaw}, exporting enters the conditional mean of productivity via two variables, of natural interest is the ability of our model to separate the direct LBE effect from indirect LFE spillovers. Using simple calculus we can show that, owing to the definition of the average \textit{peer} export orientation which excludes the export information pertaining to the $i$th firm:
\begin{align}\label{eq:lbelfeident}
\frac{\partial E[\omega_{it}|\cdot]}{\partial X_{it-1}}&=
\underbrace{\frac{\partial h(\cdot)}{\partial X_{it-1}}}_{LBE_{it}} + 
\underbrace{\frac{\partial h(\cdot)}{\partial \overline{X}_{it-1}}}_{LFE_{it}}
\times \frac{\partial \overline{X}_{it-1}}{\partial X_{it-1}}
= LBE_{it},
\end{align}
because ${\partial \overline{X}_{it-1}}\big/{\partial X_{it-1}}=\partial \sum_{j\ne i} p_{ijt}X_{jt-1} \big/\partial X_{it-1}=0$. Thus, $LBE_{it}$ is \textit{separably} identifiable. Intuitively, all observable variation in expected productivity due to a change in the firm's own export intensity is attributable to the direct learning effect because its exporting does not immediately affect the export behavior of its peers. Obviously, the separability of the two effects would be impossible if, in place of the average \textit{peer} export intensity, we would have used the total average of \textit{all} firms as oftentimes done in the literature \citep[e.g.,][]{alvarez2008exporting,jung2010}.

%Lastly, owing to the unspecified nonparametric form of the conditional mean of $\omega_{it}$ in (\ref{eq:productivitylaw}), we are able to obtain observation-specific estimates of the LBE and LFE effects thus allowing for potential cross-firm heterogeneity in the link between exporting and productivity. This also enables us to explore potential nonlinearities in the productivity effects of exporting whereby they can interact with each other as well as with the firm's own productivity.

% ------------------------------------------------------------------------------------------
% ------------------------------------------------------------------------------------------

\section{Empirical Strategy}\label{sec:empiricalframework}

Estimating the production function using least squares would result in a simultaneity bias due to the dependence of freely varying inputs (regressors in the production function) on unobserved firm productivity $\omega_{it}$ because the latter is a part of the firm's information set $\mathcal{I}_{it}$ based upon which it makes optimal input allocation decisions at $t$. This omitted variable bias is also known as a ``transmission bias" \citep{griliches1995production}. A proxy variable method proposed by \citet{olley1992dynamics} and extended by \citet{levinsohn2003estimating} tackles this endogenous problem by proxying for unobservable $\omega_{it}$ via the observable static intermediate input $M_{it}$ and then using weakly exogenous higher-order lags of inputs to instrument for endogenous freely varying inputs. Recently, this methodology has been critiqued for its lack of identification due to perfect functional dependence between freely varying inputs and self-instrumenting quasi-fixed factors \citep{ackerberg2015identification} and violation of the ``rank condition'' in instrumentation of these endogenous freely varying inputs \citep{gandhi2018identification}. As a solution,  \citet{gandhi2018identification} have suggested employing the information contained in the first-order condition for static inputs to identify both the production function and latent firm productivity. However, because their procedure is fully nonparametric, its implementation is three-stage and quite computationally burdensome, especially in its requirement to integrate the estimated static input elasticity function at each observation in order to recover the unknown production function. In this paper, we therefore rely on \citeauthor{malikovzhao2019}'s (2021) more easy-to-implement semiparametric adaptation of the \citet{gandhi2018identification}
methodology (which we modify to suit our research question) that utilizes a prespecified parametric form of the production function to derive the proxy function. This is similar to the idea pursued by \citet{doraszelski2013r}. The semiparametric adaptation can significantly ease the demand on data as well as the computational burden of estimation.

\medskip

\textsl{Identification.}\textemdash Consider the firm's optimality condition for materials. Since the intermediate input $M_{it}$ is freely varying and thus affects profits only in the current period, the firm's restricted expected profit maximization problem with respect to $M_{it}$ is as follows:
\begin{equation}\label{eq:profitmaximaztion}
\underset{M_{it}}{\max}\ P_{t}^{Y}A_{0}K_{it}^{\alpha_{K}}L_{it}^{\alpha_{L}}M_{it}^{\alpha_{M}}\exp\left\{ \omega_{it}\right\} \theta-P_{t}^{M}M_{it},
\end{equation}
where $P_{t}^{Y}$ and $P_{t}^{M}$ respectively denote the output and material input price, both of which are competitively determined. The constant $\theta$ is defined as  $\theta\equiv E\left[\exp\left\{ \eta_{it}\right\} \mid\mathcal{I}_{it}\right]$.

Taking the log-ratio of the first-order condition with respect to $M_{it}$
\begin{equation}\label{eq:firstorder}
\alpha_{M}P_{t}^{Y}A_0K_{it}^{\alpha_{K}}L_{it}^{\alpha_{L}}M_{it}^{\alpha_{M}-1}\exp\left\{ \omega_{it}\right\} \theta=P_{t}^{M}
\end{equation}
and the production function in \eqref{eq:prodfun} gives
\begin{equation}\label{eq:share}
\ln\left(S_{it}^M\right)=\ln\left(\alpha_{M}\theta\right)-\eta_{it},
\end{equation}
where $S_{it}^M\equiv\frac{P_{t}^{M}M_{it}}{P_{t}^{Y}Y_{it}}$ is
the intermediate input share of output. Thus, we can identify a composite constant $\alpha_{M}\theta$ from the unconditional moment $E\left[\eta_{it}\right]=0$, from where we have that
\begin{equation}\label{eq:firststage}
\ln\left(\alpha_{M}\theta\right)=E\left[\ln\left(S_{it}^M\right)\right].
\end{equation}

We can also separately identify the constant $\theta$ via $
\theta\equiv E\left[\exp\left\{ \eta_{it}\right\} \right]=E\left[\exp\left\{ \ln\left(\alpha_{M}\theta\right)-\ln\left(S_{it}^M\right)\right\} \right]=$ $E\left[\exp\left\{ E\left[\ln\left(S_{it}^M\right)\right]-\ln\left(S_{it}^M\right)\right\} \right]$,
with equation \eqref{eq:firststage} used to substitute for $\ln\left(\alpha_{M}\theta\right)$ in the third equality. Combining this result with \eqref{eq:firststage}, we identify the firm's material elasticity $\alpha_{M}$ as
\begin{equation}\label{eq:beta}
\alpha_{M}=\exp\left\{ E\left[\ln\left(S_{it}^M\right)\right]\right\} \slash E\left[\exp\left\{ E\left[\ln\left(S_{it}^M\right)\right]-\ln\left(S_{it}^M\right)\right\} \right],
\end{equation}
where it is a unique function of the first moments of data.

To identify the rest of production function as well as latent firm productivity, we take the log of \eqref{eq:prodfun} on both sides to obtain
\begin{equation}\label{eq:logprodfun}
y_{it}=\alpha_{0}+\alpha_{K}k_{it}+\alpha_{L}l_{it}+\alpha_{M}m_{it}+\omega_{it}+\eta_{it},
\end{equation}
where $\alpha_{0}\equiv \ln A_{0}$; and the lower-case variables correspond to the log form of the respective upper-case variables. Exploiting the Markov process
of $\omega_{it}$ in \eqref{eq:productivitylaw} and bringing the already identified material elasticity $\alpha_{M}$ to the left-hand side, we rewrite \eqref{eq:logprodfun} as follows:
\begin{equation}\label{eq:almostfinal}
y_{it}^{*}=\alpha_{K}k_{it}+\alpha_{L}l_{it}+g\left(\omega_{it-1},X_{it-1},\overline{X}_{it-1}\right)+\zeta_{it}+\eta_{it},
\end{equation}
where $y_{it}^{*}=y_{it}-\alpha_{M}m_{it}$ is fully identified and can be treated as an observable, and $g\left(\cdot\right)\equiv h(\cdot)+\alpha_{0}$ is
of unknown functional form.

Next, from equation \eqref{eq:firstorder} we derive the explicit form of the conditional demand function for $M_{it}$, which we then invert to proxy for the unobservable scalar $\omega_{it}$ in \eqref{eq:almostfinal} in the spirit of material-based proxy estimators:
\begin{equation}\label{eq:final}
y_{it}^{*}=\alpha_{K}k_{it}+\alpha_{L}l_{it}+g\left(\left[m_{it-1}^{*}-\alpha_{K}k_{it-1}-\alpha_{L}l_{it-1}\right],X_{it-1},\overline{X}_{it-1}\right)+\zeta_{it}+\eta_{it},
\end{equation}
where $m_{it-1}^{*}=\ln\left({P_{t-1}^{M}}\big/{P_{t-1}^{Y}}\right)-\ln\left(\alpha_{M}\theta\right)-\left(\alpha_{M}-1\right)m_{it-1}$ is also fully identified and treated as an observable. Since all regressors appearing in \eqref{eq:final} including $k_{it}$, $l_{it}$, $k_{it-1}$, $l_{it-1}$, $m_{it-1}^*(m_{it-1})$, $X_{it-1}$ and $\overline{X}_{it-1}$ are weakly exogenous based on our structural assumptions, there is no endogenous covariate on the right-hand side of the equation. That is,
\begin{equation}\label{eq:exo}
E\left[\zeta_{it}+\eta_{it}\bigl\vert k_{it},l_{it},k_{it-1},l_{it-1},m_{it-1}^*(m_{it-1}),X_{it-1},\overline{X}_{it-1}\right]=0,
\end{equation}
and the equation \eqref{eq:final} is identified. 

From \eqref{eq:final}--\eqref{eq:exo}, it is obvious that, if there were indeed non-zero export spillovers and we had failed to account for them in the firm's productivity evolution process, then $\overline{X}_{it-1}$ would have been omitted from the proxy function $g(\cdot)$ in \eqref{eq:final} and, consequently, been absorbed, along with its interactions with other arguments of the proxy, into the error term. In the latter case, the error term would then contain variation from the firm's quasi-fixed inputs, its own export intensity as well as the average export orientation of its peers. Generally, these all would be correlated with quasi-fixed inputs and the export variable included as regressors thus violating the weak exogeneity condition, and the model would be unidentified due to the omitted variable bias. This  highlights the importance of embedding the external spillover channel into the analytical framework explicitly. 

Lastly, we can recover latent firm productivity up to a constant using the identified production-function parameters and the productivity shock: $\omega_{it}+\alpha_{0}= y_{it}-\alpha_{K}k_{it}-\alpha_{L}l_{it}-\alpha_{M}m_{it}-\eta_{it}$.

\medskip

\textsl{Estimation procedure.}\textemdash The estimation is simple and involves a two-stage procedure. In the first stage, we estimate $\alpha_{M}$ via a sample counterpart of \eqref{eq:beta} constructed using sample averages computed from the raw data on material share:
\begin{equation*}
\widehat{\alpha}_{M}=\exp\left\{ \frac{1}{nT}\sum_{i}\sum_{t}\ln\left(S_{it}^M\right)\right\} \Bigg/\left[\frac{1}{nT}\sum_{i}\sum_{t}\exp\left\{\left[ \frac{1}{nT}\sum_{i}\sum_{t}\ln\left(S_{it}^M\right)\right]-\ln\left(S_{it}^M\right)\right\} \right].
\end{equation*}

As a by-product, we also have $\ln\widehat{\left(\alpha_{M}\theta\right)}= \frac{1}{nT}\sum_{i}\sum_{t}\ln\left(S_{it}^M\right)$ and $\widehat{\eta}_{it}=\ln\widehat{\left(\alpha_{M}\theta\right)}-\ln\left(S_{it}^M\right)$. With these estimates in hand, we then construct estimates of $y_{it}^{*}$ and
$m_{it}^{*}$ as $\widehat{y}_{it}^{*}=y_{it}-\widehat{\alpha}_{M}m_{it}$ and
$\widehat{m}_{it-1}^{*}=\ln\left(\frac{P_{t-1}^{M}}{P_{t-1}^{Y}}\right)-\ln\widehat{\left(\alpha_{M}\theta\right)}-\left(\widehat{\alpha}_{M}-1\right)m_{it-1}$,
respectively.

The second-stage estimation requires the choice of an approximator for the unknown $g(\cdot)$. We use the popular second-order polynomial
sieves \citep[e.g.,][]{gandhi2018identification}. Specifically, we approximate $g\left(\cdot\right)$ as follows: \small
\begin{align}
g\left(\cdot\right) & \approx\left(1,W_{it-1}\left(\mathbf{\boldsymbol\alpha}\right),W_{it-1}^{2}\left(\boldsymbol{\alpha} \right),X_{it-1},X_{it-1}^{2},\overline{X}_{it-1},\overline{X}_{it-1}^{2},W_{it-1}\left(\boldsymbol{\alpha} \right)X_{it-1},W_{it-1}\left(\boldsymbol{\alpha}\right)\overline{X}_{it-1},X_{it-1}\overline{X}_{it-1}\right)'\boldsymbol{\gamma}  \equiv\boldsymbol{\lambda}_{it}\left(\boldsymbol{\alpha}\right)'\boldsymbol{\gamma}, \label{eq:omegaapprox}
\end{align} \normalsize
where $\boldsymbol{\alpha}=\left(\alpha_{K},\alpha_{L}\right)'$, $W_{it-1}\left(\boldsymbol{\alpha} \right)=\widehat{m}_{it-1}^{*}-\alpha_{K}k_{it-1}-\alpha_{L}l_{it-1}$, and $\boldsymbol{\gamma}$ is the unknown parameter vector.

We estimate \eqref{eq:final} using a nonlinear
least squares method to obtain the second-stage estimates of $(\alpha_{K},\alpha_{L})'$ and $\boldsymbol{\gamma}$:
\begin{equation*}
\underset{\alpha_{K},\alpha_{L},\boldsymbol{\gamma}}{\min}\ \sum_{i}\sum_{t}\left(\widehat{y}_{it}^{*}-\alpha_{K}k_{it}-\alpha_{L}l_{it}-\boldsymbol{\lambda}_{it}\left(\alpha_{K},\alpha_{L}\right)'\boldsymbol{\gamma} \right)^{2}.
\end{equation*}

With the estimated production-function parameters in hand, we can compute the productivity effects via $\widehat{LBE}_{it}={\partial\widehat{g}\left(\cdot\right)}\big/{\partial X_{it-1}}$ and $\widehat{LFE}_{it}={\partial\widehat{g}\left(\cdot\right)}\big/{\partial\overline{X}_{it-1}}$, where $\widehat{g}\left(\cdot\right)=\boldsymbol{\lambda}_{it}\left(\widehat{\boldsymbol\alpha}\right)'\widehat{\boldsymbol{\gamma}}$.\footnote{Note that the definitions of the LBE and LFE effects are based on the gradients of $h(\cdot)$ but, since $g(\cdot)$ and $h(\cdot)$ differ only by an additive constant, their gradients are equal.}
We also recover $\omega_{it}$ up to a constant via
$\widehat{\omega_{it}+\alpha_{0}}=y_{it}-\widehat\alpha_{K}{k}_{it}-\widehat\alpha_{L}{l}_{it}-\widehat\alpha_{M}{m}_{it}-\widehat{\eta}_{it}$.

For statistical inference, we employ accelerated biased-corrected percentile bootstrap confidence intervals. Details are provided in Appendix \ref{sec:appx_boot}.

% ------------------------------------------------------------------------------------------
% ------------------------------------------------------------------------------------------

\section{Data}\label{sec:data}

Our data come from the Encuesta Nacional Industrial Anual (ENIA), a national industrial survey, conducted by the Chilean National Institute of Statistics annually. The sample period runs from 1995 to 2007. Manufacturing plants are classified into 22 industry groups according to the 2-digit International Standard Industry Classification (ISIC) code. The dataset contains information on plants from 13 regions including Tarapac\'a, Antofagasta, Atacama, Coquimbo, Valpara\'iso, Libertador Gral.~Bernardo O'Higgins, Maule, Biob\'io, La Araucan\'ia, Los Lagos, Ais\'en del Gral.~Carlos Ib\'a\~{n}ez del Campo, Magallanes and Chilean Antarctica, and the Santiago Metropolitan region. Though each observation represents a plant rather than a firm, single-plant establishments account for over 90\% of total units \citep[also see][]{pavcnik2002trade}.

The total output is defined as total revenues from the sale of products and work done and, as such, our analysis is carried out using a revenue-based productivity measure (TFPR). As in \citet{de2013detecting}, the measured TFPR will capture export-driven changes in both the cost and demand factors. Hence, a positive LBE and/or LFE effect can be interpreted as working through either a decline in production cost or an increase in demand, or both. For more on the latter point, see Appendix \ref{sec:appx_a}.

Capital is the fixed assets balance for buildings, machinery and vehicles at the end of a period. Materials are defined as the total expenditure on intermediate inputs consisting of raw materials and other intermediates. These three variables are measured in hundred thousands of pesos, and we deflate them using price deflators at the 4-digit ISIC level. We measure labor using the total number of people working at the plant. We drop observations that contain missing or negative values for these variables and exclude extreme outliers lying outside the interval between the 1th and 99th percentiles of these four variables. In the end, our sample consists of 8,353 manufacturing plants with a total of 47,622 observations.

Export intensity is calculated as the nominal proportion of the firm's exports in its total sales, ranging from 0 to 1 by construction. Out of all plants, exporters are 20.6\% of the sample. As discussed earlier, we measure each plant's exposure to exporters using the average export intensity of its peers, excluding the ``recipient'' plant itself. Peers are identified as operating in the same one of 13 regions and the same 2-digit industry within the same period (a total of 186 peer groups), but we remove the geographic limits in one of the robustness checks. The exposure variable also ranges between 0 and 1 by construction.  Table \ref{tab:data} in Appendix \ref{sec:appx_data} provides the summary statistics for our data.

% ------------------------------------------------------------------------------------------
% ------------------------------------------------------------------------------------------

\section{Results}\label{sec:results}

Our primary interest is in the estimates of productivity effects of exporting. Owing to a nonparametric form of the conditional mean of $\omega_{it}$, we obtain observation-specific estimates of the LBE and LFE effects. Since these effects are defined as the gradients of expected future firm log-productivity with respect to export intensity or the peer average thereof, both of which are proportions, the reported $LBE_{it}$ ($LFE_{it}$) is a semi-elasticity measuring percentage changes in productivity per unit percentage point change in the firm's export intensity (the average peer export orientation of the local industry). Lastly, both effects measure ``short-run'' productivity improvements per annum, which however can accumulate over time owing to the persistent nature of productivity evolution.

\iffalse
% -----------------------------------
\begin{table}[t]
	\caption{Production Function Parameter Estimates}\label{tab:productionestimates}
	\centering
	\begin{threeparttable}
		\begin{tabular}{l ccc}
			\toprule[1pt]
			Parameter & Point Estimate & Lower Bound & Upper Bound\\
			\midrule
			
			Capital Elasticity  & 0.214 & 0.204 & 0.226 \\
			Labor Elasticity    & 0.456 & 0.435 & 0.476 \\
			Material Elasticity & 0.288 & 0.284 & 0.291 \\[3pt]
			Scale Elasticity    & 0.957 & 0.938 & 0.975\\
			
			\midrule
			\multicolumn{4}{p{12.3cm}}{\textit{Notes}: Reported are the input elasticity estimates along with their two-sided 95\% lower and upper confidence bounds. Scale elasticity is the sum of capital, labor and material elasticities.}\\
			\bottomrule[1pt]
		\end{tabular}
		
	\end{threeparttable}
\end{table}
% ------------------------------
\fi 

\subsection{Exporter productivity differential}

As a preliminary analysis, we first examine the overall cross-firm differential in productivity $\omega_{it}$ across exporters and non-exporters in the Chilean manufacturing sector.\footnote{The associated production function parameter estimates are $\widehat{\alpha}_K=0.21$, $\widehat{\alpha}_L=0.46$ and $\widehat{\alpha}_M=0.29$. Scale elasticity, defined as the sum of all three input elasticities, is 0.96 with the corresponding 95\% bootstrap confidence interval of (0.938, 0.975), indicating the decreasing returns to scale, consistent with the profit-maximizing behavior. Allowing for differences in estimators and sample periods, our estimates are in the same ballpark as the input elasticity and returns to scale values reported by other studies that used Chilean data.} The mean estimate of the (log)-productivity differential between exporters and non-exporters is 0.28 with the corresponding 95\% bootstrap percentile confidence interval of  (0.26, 0.31), which indicates that, on average, exporters are more productive than non-exporters by roughly 28\%. 

\begin{figure}[t]
	\centering
	\includegraphics[scale=0.55]{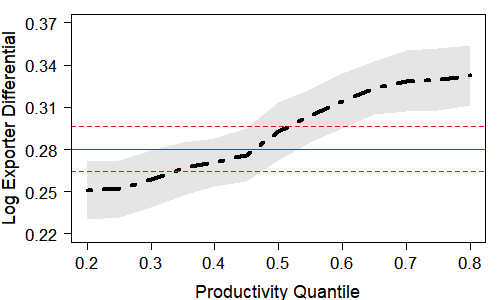}
	\caption{Exporter Productivity Differential Estimates across Productivity Quantiles\newline {\small [\textit{Notes}: Shaded bands correspond to 95\% confidence intervals. Solid horizontal lines correspond to the productivity differentials estimated at the conditional mean, with their respective 95\% confidence intervals shown using dashed lines.]} }\label{fig:wexport}
\end{figure}

We further investigate if a plant's exporter status commands a productivity premium of varying magnitude and significance at different points in the productivity distribution. We do so by estimating a simple quantile regression: $Q_{\tau}\left[\omega_{it}|\cdot\right]= \beta_{0,\tau}+\beta_{1,\tau}1(X_{it}>0)\ \text{for}\ \tau\in(0,1)$. Employing a quantile regression enables us to explore potential distribution heterogeneity in the exporter productivity differential. We estimate this model for the quantile index $\tau$ taking values from 0.2 to 0.8 (with 0.05 increments) thereby focusing on the central portion of the productivity distribution. Figure \ref{fig:wexport} plots the quantile regression estimates of the $\beta_{1,\tau}$ coefficient (the exporter productivity differential) against $\tau$, along with the 95\% confidence intervals. %The solid horizontal line corresponds to the productivity differential estimated at the conditional mean. 
The quantile productivity differentials are all significantly positive and increasing with the quantile of $\omega$, indicating that the productivity divergence between exporters and non-exporters is more prominent magnitude-wise among more productive firms.\footnote{Including controls for the plant size (proxied by the number of employees) as well as the region and year effects produces similar findings.} 

% -----------------------------------
% -----------------------------------

% -----------------------------------
\begin{figure}[t]
	\centering
	\includegraphics[scale=0.55]{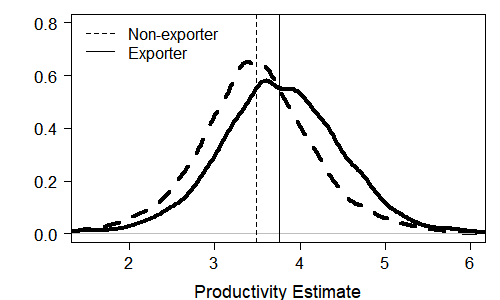}
	\caption{Distributions of log-Productivity by the Exporter Status
	\newline {\small [\textit{Notes}: Vertical lines show respective sample means.]}}\label{fig:productivity}
\end{figure} 
% -----------------------------------

For a more holistic look at the productivity differential between exporters and non-exporters, we also plot the kernel density of firm log-productivity for exporters and non-exporters, as shown in Figure \ref{fig:productivity}. This allows us to compare distributions of productivity estimates as opposed to merely focusing on marginal moments or quantiles. The figure indicates that exporters appear to enjoy a productivity premium over non-exporters distribution-wise. To support this visual evidence, we do a formal test to check if exporters are more productive than non-exporters along the entire distribution of productivity. We utilize a generalization of the Kolmogorov-Smirnov test proposed by \citet{linton2005consistent} to test the (first-order) stochastic dominance of exporters'
productivity over non-exporters'. This test permits the variables to be estimated latent quantities as opposed to observables from the data and to also share dependence (in our case, the dependence is due to their construction using the same set of parameter estimates). 
%Specifically, let $G_{1}\left(\omega\right)$ and $G_{0}\left(\omega\right)$ denote the cumulative distribution functions of productivity $\omega\in\Omega$ for exporters and non-exporters, respectively. We then construct the null hypothesis that non-exporters' productivity is stochastically dominated by that of exporters as follows:
%\begin{equation}
%H_{0}:\ \underset{\omega\in\Omega}{\sup}\left[G_{1}\left(\omega\right)-G_{0}\left(\omega\right)\right]\leq0\quad \textrm{vs}.\quad H_{1}:\ \underset{\omega\in\Omega}{\sup}\left[G_{1}\left(\omega\right)-G_{0}\left(\omega\right)\right]>0,\notag
%\end{equation}
%with the corresponding test statistic defined as $\mathscr{D}=\max_{1\leq j\leq\left(n_{1}+n_{0}\right)}\sqrt{\frac{n_{1}n_{0}}{n_{1}+n_{0}}}\left[G_{1,n_{1}}\left(\widehat{\omega}_{j}\right)-G_{0,n_{0}}\left(\widehat{\omega}_{j}\right)\right]$, where $G_{s,n_{s}}$ is the empirical distribution function of the estimated (log) productivity $\omega$ for the $s$th category of plants of the sample size $n_{s}$, with $s=\{ 0, 1\}$. 
Employing the sub-sampling procedure from \citet{linton2005consistent}, we obtain a $p$-value for the test statistic of 0.7789 %\footnote{We use $r_n$ equidistant sub-sample sizes $B_{n}=\left\{ b_{1},\cdots,b_{r}\right\} $, where $b_{1}=\left[\log\log n\right]$, $b_{r_n}=\left[n/\log\log n\right]$, and the number of unique sub-sample sizes is $r=199$. For each $b$, we get a $p$-value. The reported is the mean of these $p$-values.} 
and, thus, comfortably fail to reject the null hypothesis that non-exporters' productivity is stochastically dominated by that of exporters. 

Altogether, we can conclude that exporters enjoy a statistically significant productivity premium over non-exporters along the entire distribution of firm productivity. 

\subsection{Learning by exporting and from exporters}

The exporter productivity premium can reflect both the learning and self-selection effects. To zoom in on the within-firm evidence of productivity-enhancing \textit{learning} effects of exporting and to test if the firm's past export experience as well as the experience of its peers impact its future productivity, we use $LBE_{it}={\partial E\big[ \omega_{it}|\omega_{it-1},X_{it-1},\overline{X}_{it-1}\big]}\big/{\partial X_{it-1}}$ and $LFE_{it}={\partial E\big[ \omega_{it}|\omega_{it-1},X_{it-1},\overline{X}_{it-1}\big]}\big/{\partial\overline{X}_{it-1}}$ which are conditioned on the firm's past productivity level $\omega_{it-1}$ thereby, as discussed in Appendix \ref{sec:appx_selection}, allowing us to account for self-selection into exporting based on an \textit{a priori} higher productivity. In effect, both LBE and LFE measure a differential in mean future productivity identified from the difference in current productivity between firms/peers with \textit{different} export experiences, holding their input use fixed. Table \ref{tab:productivityeffect}  reports a summary of point estimates of these LBE and LFE effects for all firms as well as for exporters and non-exporters only. We also test for the statistical significance of these effects at \textit{each} observation. The shares of observations for which each of the two productivity effects of exporting is significant at the 5\% level are provided in the far right column of the table.

The LBE effect is estimated to average at 0.36 for the entire sample, indicating that a percentage point increase in the firm's own export intensity raises its future productivity by 0.36\%. The point estimates are between 0.32 and 0.45 within the inter-quartile range. Overall, the LBE productivity effect is statistically significant for 93\% of firms in our data set. We also document notable differences in the magnitude and prevalence of LBE across exporters and non-exporters. For exporters ($X_{it-1}>0$), the mean LBE effect is significant at 0.158, albeit the observation-specific point estimates are statistically non-zero only for 68\% of the exporting firms. But the evidence is more ubiquitous among non-exporters. While the latter category of firms does not actually export, our methodology still produces an estimate of $LBE_{it}={\partial E\big[ \omega_{it}|\cdot\big]}\big/{\partial X_{it-1}}$ for them which, in this case, is evaluated at $X_{it-1}=0$. Essentially, these estimates of the LBE effect are ``counterfactual'' and provide a measurement of how much non-exporters' productivity would change if they were to begin exporting (i.e., marginally increase their export intensity from zero to a positive value).\footnote{We can estimate the LBE effect for \textit{non}-exporters directly by the virtue of algebra of the regression and, as pointed out by a referee, this estimand is similar in character to the concept of a treatment effect on the non-treated. When estimating the second-stage regression \eqref{eq:final} we approximate the nonlinear productivity process $g(\omega_{it-1} X_{it-1},\overline{X}_{it-1})$ via quadratic polynomial given in \eqref{eq:omegaapprox} and thus obtain heterogeneous estimates of the gradient \textit{function} measuring LBE, $\widehat{\partial g(\cdot)/\partial X_{it-1}}= \widehat{\lambda}_{x}+0.5\widehat{\lambda}_{xx}X_{it-1}+ \widehat{\lambda}_{x\omega}\widehat{\omega}_{it-1}+\widehat{\lambda}_{x\bar{x}}\overline{X}_{it-1}$ where $\widehat{\lambda}$'s are the estimated coefficients. It can be evaluated at each observation, including for the firms that export ($X_{it-1}\ne0$) and do not export ($X_{it-1}=0$) in period $t-1$. The distinction between the LBE effects for exporters and non-exporters has been typically under-emphasized in prior studies many of which relate firm productivity and $X$ linearly, obtaining a singular estimate of the LBE effect. Owing to the homogeneity of treatment effects implied by linearity, such an LBE estimate applies to both exporters and non-exporters. We however allow for heterogeneity in the effects. } The average LBE estimate for non-exporters is 0.418 and significant. In fact, the point estimates of the LBE effect are statistically significant for virtually all non-exporters (99\%). Magnitude-wise, the average effect size for non-exporters is about 2.6 times larger than that for active exporters. These findings suggest that the bulk of a productivity boost attributable to (internal) learning from exporting is realized upon the domestic firm's entry into in export markets as it gains access to new technology and business practices, and the effectiveness of LBE is diminishing as the firm further specializes in exporting.\footnote{The diminishing productivity return to the own export experience is also corroborated by the estimates of the LBE function which we discuss later.} This initial effect likely captures the productivity-boosting effects of various export-related investments that new exporters normally undertake. Namely, the decisions to start exporting tend to go together with other firm-level actions that can enhance productivity such as the adoption of new/upgraded technologies, quality upgrading or R\&D spending \citep[e.g., see][]{verhoogen2008trade,aw2011r,bustos2011trade}. Due to the lack of richer data, we are unable to unbundle these individual effects, and our LBE estimates ought to be interpreted as measuring their composite export-focused effect attributable to multiple channels.

% -----------------------------------
\begin{table}[t]
	\caption{The LBE and LFE Productivity Effect Estimates}\label{tab:productivityeffect}
	\centering\small
	\makebox[\linewidth]{
		\begin{tabular}{lcccc|c}
			\toprule[1pt]
			& \multicolumn{4}{c}{\it Point Estimates} & Stat.~Signif. \\
			& Mean & 1st Qu. & Median & 3rd Qu. & (\% Obs.) \tabularnewline
			\midrule
			\multicolumn{6}{c}{\textbf{---Learning by Exporting---}} \\
			All Firms        & 0.363         & 0.323          & 0.390          & 0.451         & 92.7 \\
			& (0.189, 0.555) & (0.139, 0.531)  & (0.207, 0.603) & (0.250, 0.675)  &      \\
			Exporters     & 0.158         &  --0.000              & 0.257         & 0.353         & 68.3 \\
			& (0.068, 0.254) & (--0.072, 0.067) & (0.117, 0.419) & (0.184, 0.546) &      \\
			Non-exporters & 0.418         & 0.356          & 0.408         & 0.465         & 99.2 \\
			& (0.234, 0.643) & (0.170, 0.572)   & (0.220, 0.628)  & (0.259, 0.695) &      \\[3pt]
			
			\multicolumn{6}{c}{\textbf{---Learning from Exporters---}} \\
			All Firms         & 0.324         & 0.141          & 0.296         & 0.457         & 68.5 \\
			& (0.116, 0.545) & (--0.059, 0.363) & (0.069, 0.520)  & (0.166, 0.729) &      \\
			Exporters     & 0.508         & 0.182          & 0.399         & 0.778         & 73.9 \\
			& (0.302, 0.76)  & (--0.013, 0.432) & (0.189, 0.618) & (0.508, 1.148) &      \\
			Non-exporters & 0.275         & 0.132          & 0.28          & 0.42          & 67.1 \\
			& (0.027, 0.489) & (--0.088, 0.34)  & (0.027, 0.497) & (0.124, 0.701) &     \\
			\midrule
			\multicolumn{6}{p{14.4cm}}{\small \textit{Notes}: Reported is a summary of point estimates of the LBE and LFE effects tabulated by the firm's exporter status, with 95\% bootstrap percentile confidence intervals in parentheses. The far right column reports the share of (sub)sample for which the observation-specific estimates are statistically significant at the 5\% level.}\\
			\bottomrule[1pt]
		\end{tabular}
	}	
\end{table}
% -----------------------------------

On average, the size of the LFE effect is on par with that of LBE: a pooled mean estimate of the LFE effect is statistically significant at 0.32 (vs.~0.36 for LBE) whereby a percentage point increase in the average peer export intensity within a domestic firm's local industry boosts its future productivity by 0.32\%. It may perhaps surprise the reader that the indirect cross-firm learning channel is estimated to boost firm productivity by almost as much as the LBE occurring internally. Of note, however, is that these two effects are not commensurable with one another because, by construction, they capture impacts of qualitatively different ``treatments.'' While $LBE_{it}$ measures the productivity effect of a unit change in \textit{one} firm's (own) export intensity $X_{it-1}$, $LFE_{it}$ measures the effect of a unit change in the average peer export intensity $n_{it-1}^{-1}\sum_{j\in \mathcal{L}_{it-1}} X_{jt-1}$ of the entire local industry, a change that is equivalent to an increase in the export orientation of \textit{all} peers by a unit. As such, the scale of two treatments is vastly different. To wit, the LFE effect quantifies a \textit{total} indirect/spillover effect on firm $i$'s productivity at time $t$ from all its peers of whom, at time $t-1$, it had $n_{it-1}$. The average spillover effect on future productivity of firm $i$ from a percentage point increase in a \textit{single} peer's $X_{jt-1}$ is a significantly more modest 0.013\%. In what follows, we continue focusing on the (total) LFE effect that aggregates spillovers from all peers\textemdash over the average spillover across only a pair of firms\textemdash because its effect size is a more policy-relevant quantity.

We find that the LFE effect is significantly non-zero only for 69\% of recipient firms, thus suggesting that the indirect cross-firm productivity-boosting effect of exporting is less prevalent in Chilean manufacturing than the direct effect taking place within a firm (recall, 93\%). Of interest, we also document that the LFE effect is stronger for firms who also export themselves. Namely, the average estimate of the LFE effect for exporter firms is estimated at 0.508, whereas the corresponding estimate for non-exporters is half that at 0.275. In addition, non-zero learning from exporters is also more prevalent for exporters (74\%) than it is for non-exporters (67\%). The empirical evidence therefore suggests that exporters benefit from the exposure to peer exporters in their local industry more than non-exporters. Plausibly, there may be complementarities between internal/direct and external/indirect learning from export experiences. This is reasonable because challenges associated with engagements in the foreign market can make exporters more motivated and pressured than their fully domestically-oriented non-exporting peers to improve further and more intensely.

It is important to re-emphasize that, given an inherently ``black box'' nature of production functions and the residual-based definition of TFP,  \textit{magnitudes} of LFE point estimates ought to be taken with caution and, probably, not at their face value. As noted earlier, interpretation of the LFE measure based on spatio-industrial proximity of exporting peers as capturing, specifically, the spillover effects of exporting requires that we assume away all other plausible channels of cross-firm interactions that may also spur productivity spillovers, including spatial agglomeration, R\&D and FDI externalities, spillovers along supply chains, etc. These other factors are expected to correlate with exporting behavior, and the LFE effect that we measure may partly reflect spillovers that they induce across firms as well. Unless all plausible internal and external productivity modifiers are controlled for when modeling firm productivity evolution, measured LFE effects will expectedly be crude. This is an obvious limitation of our analysis but also happens to be the case for any other study that has ever sought to measure productivity spillovers. (The same arguments can also be made about LBE.) Consequently, we mainly pursue a simpler,  more feasible objective of testing for \textit{non-zero} export-driven spillover effects on productivity. That is, our focus is on the ``extensive margin,'' i.e., whether productivity spillovers from peer exporters occur in general, as opposed to how big these effects are. To this end, we find that the productivity externalities from exporting peers are significant in two-thirds of plants.

\medskip

\textsl{Cumulative Long-Run Effects.}\textemdash As noted earlier, the reported point estimates of the LBE and LFE effects are short-run and do not account for dynamic effects over time. Obviously, owing to the persistence of productivity, the cumulative implications of exporting for domestic firms' productivity in the long run will be more sizable. Because the estimated productivity process is nonlinear, the computation of such cumulative effects is, however, not clear-cut. To roughly size up the economic significance of LBE and LFE for productivity in the domestic industry in the long-run equilibrium, we perform the following back-of-envelope calculation. Approximating $h(\cdot)$ using a linear expansion around the point such that its evaluated gradients are equal to the mean estimates from our semiparametric model, i.e., 
\begin{equation*}
	\omega_{it}\approx\text{const}+ 0.482\omega_{it-1}+0.363X_{it-1}+0.324\overline{X}_{it-1}+\text{error}_{it},
\end{equation*}
under the temporal stationarity of log-productivity we have that the long-run LBE and LFE effects are respectively equal to $0.363/(1-0.482)=0.701$ and $0.324/(1-0.482)=0.627$. Thus, in the long-run equilibrium a permanent 10 percentage point increase in mean export intensity of \textit{all} firms in a local industry is roughly estimated to produce a sizable 13.3\% boost to the average productivity of domestic firms.\footnote{The total long-run effect on $E[\omega]$ of a 10 percentage point increase in $E[X]$ is $\text{d} E[\omega]/\text{d} E[X]\times 10=(\partial E[\omega]/\partial E[X]+\partial E[\omega]/\partial E[\overline{X}]\times\partial E[\overline{X}]/\partial E[X])\times10=(0.701+0.627)\times 10=13.3$, where we have made use of $\partial E[\overline{X}]/\partial E[X]=1$.} This estimate incorporates both direct LBE and indirect LFE channels as well as the temporal multiplier effect due to persistence over time. Whether occurring through technology/efficiency-driven cost savings or an increase in demand (or both), this improvement in the long-run average TFPR, $E[\omega]$, associated with an industry's 10\% pivot towards more exporting is roughly equivalent to $13.3\%\times E[Y]=3,680$ thousands pesos more in revenues for an average domestic firm, holding the input usage fixed.

\medskip

\textsl{Heterogeneity and Nonlinearity.}\textemdash Although we find that average effect sizes of LBE and LFE are commensurate, individual firms are highly heterogeneous across many dimensions including their productivity, the degree of their export orientation as well as the intensity of their exposure to other exporters in the industry. It is therefore of interest to investigate if these characteristics influence the effect size of their internal and external learning from exporting. 

Recall that we obtain the estimate of the productivity effects of exporting via $\widehat{LBE}_{it}={\partial\widehat{g}\left(\cdot\right)}\big/{\partial X_{it-1}}$ and $\widehat{LFE}_{it}={\partial\widehat{g}\left(\cdot\right)}\big/{\partial \overline{X}_{it-1}}$,
where we estimate $g\big(\omega_{it-1},X_{it-1}, \overline{X}_{it-1}\big)$ using the second-order polynomial sieve approximation given in \eqref{eq:omegaapprox}. Thus, by derivation, both $\widehat{LBE}_{it}$ and $\widehat{LFE}_{it}$ are the estimated linear \textit{functions} of the ``determinants'' of firm productivity $\big(\omega_{it-1},X_{it-1}, \overline{X}_{it-1}\big)'$.\footnote{Concretely, under the quadratic polynomial approximation of $g(\cdot)$ in \eqref{eq:omegaapprox}, the estimated gradient functions are $\widehat{LBE}_{it}\equiv{\partial \widehat{g}(\cdot)/\partial X_{it-1}}= \widehat{\lambda}_{x}+0.5\widehat{\lambda}_{xx}X_{it-1}+ \widehat{\lambda}_{x\omega}\widehat{\omega}_{it-1}+\widehat{\lambda}_{x\bar{x}}\overline{X}_{it-1}$ and $\widehat{LFE}_{it}\equiv{\partial \widehat{g}(\cdot)/\partial \overline{X}_{it-1}}= \widehat{\lambda}_{\bar{x}}+ 0.5\widehat{\lambda}_{\bar{x}\bar{x}}\overline{X}_{it-1}+ \widehat{\lambda}_{\bar{x}\omega}\widehat{\omega}_{it-1}+\widehat{\lambda}_{x\bar{x}}X_{it-1}$ where $\widehat{\lambda}$'s are the estimated polynomial coefficients. Table \ref{tab:effectpath} reports the coefficients on regressors in these functions.}
Table \ref{tab:effectpath} reports the estimates of parameters on these three variables for LBE and LFE. 

% -----------------------------------
\begin{table}[t]
	\caption{Estimates of the LBE and LFE Functions}\label{tab:effectpath}
	\centering	\small
	\begin{threeparttable}
		\begin{tabular}{lcc}
			\toprule[1pt]
			& LBE & LFE\tabularnewline
			\midrule
			$\omega_{it-1}\qquad $ &	--0.113          & --0.346          \\
			&	(--0.212, --0.039)   & (--0.575, --0.134)   \\
			$X_{it-1}$ &	--0.970           & 1.210            \\
			&	(--1.454, --0.435) & (0.583, 2.008)   \\
			$\overline{X}_{it-1}$ &	1.210            & 0.220           \\
			&	(0.583, 2.008)   & (--0.959, 1.377)  \\
			\midrule
			\multicolumn{3}{p{7.2cm}}{\small \textit{Notes}: Reported are the parameter estimates for the LBE and LFE functions derived from the polynomial approximation of the conditional mean of $\omega_{it}$, along with the 95\% bootstrap percentile confidence intervals in parentheses. }\\
			\bottomrule[1pt]
		\end{tabular}
		
	\end{threeparttable}
	
\end{table} 
% -----------------------------------

The coefficient estimates on $\omega_{it-1}$ for both LBE and LFE are significantly negative, indicating that the magnitude of these effects declines as firms get more productive. Thus, less productive plants benefit more from export-driven learning, be it an internal or external channel. This finding suggests that plants that are already highly productive have less to gain from export-driven learning as well as less to learn from their exporting peers, presumably because they're much closer to the technology/knowledge frontier compared to less productive domestic firms. We also find that a firm's own export intensity $X_{it-1}$ has a significantly negative effect on learning by exporting but a significantly positive effect on learning from exporters, indicating that less export-oriented plants improve more via learning by exporting and less via learning from exporters (and vice versa). Basically, this is indicative of the diminishing productivity return to the own export experience: an increase in the degree of a firm's export orientation enhances its productivity at a decreasing rate. But we do not find such a pattern for the LFE effect. On the contrary, the more export-oriented the plant is, the higher the cross-firm export-driven productivity spillovers are, which buttresses our earlier discussion of potential complementarities between exporting and external learning from Table \ref{tab:productivityeffect}. Lastly, the results in Table \ref{tab:effectpath} suggest that the average of peer export orientation in local industry $\overline{X}_{it-1}$ has a significantly positive influence on learning by exporting, indicating that a greater exposure to exporters helps plants absorb productivity improvements from their own export behavior. At the same time, we find no significant effect of $\overline{X}_{it-1}$ on LFE effect size, indicating that the average export participation in the industry improves plant productivity at a constant rate.

\medskip

\textsl{Robustness Analysis.}\textemdash We also assess the robustness of our main empirical finding of significant LBE and LFE effects to the choice of how we measure (\textit{i}) export experience of firms and (\textit{ii}) their exposure to export experiences of their peers. Specifically, we re-estimate our model using a binary export-status variable $1(X_{it}>0)$ as well as redefine $\overline{X}_{it}$ by no longer restricting the pool of a firm's peers to the same geographic region. The latter is motivated by the plausible arguments that, unlike pooling of workers, the knowledge diffusion mechanism need not be local and may spill widely across space. These results are summarized in Table \ref{tab:productivityeffect_robust}. Two observations are in order here. First, using a discrete measure of export experience yields smaller effect sizes for both LBE and LFE. Second, expanding the group of peer firms to include all regions, thereby covering a wider spillover pool, results in significantly larger estimates of the external LFE effect while expectedly having no notable impact on the internal LBE effect.\footnote{This is so despite that the indirect LFE effect \textit{per} a firm's each peer actually gets diluted; see tables in Appendix \ref{sec:appx_addresults}.} The latter indirectly lends support to the hypothesis that export spillover effects may have a broad geographical scope. By restricting the extent of spillovers to the same region in our main specification, we also restrict the reach of cross-firm export externalities. 
	
% -----------------------------------
\begin{table}[t]
	\caption{Robustness to Alternative Measures of Exporting and of the Exposure to Exporters}\label{tab:productivityeffect_robust}
	\centering\small
	\makebox[\linewidth]{
		\begin{tabular}{l cccc}
			\toprule[1pt]
			& (I) & (II) & (III) & (IV) \\
			\midrule
			LBE & 0.363  		&  0.379 		& 0.147 		& 0.151 		 \\
				& (0.189, 0.555)& (0.207, 0.575)& (0.117, 0.176)& (0.121, 0.179) \\
				& [92.7\%] 		& [89.4\%] 		& [99.6\%] 		& [99.6\%] \\[5pt]
				
			LFE & 0.324  		& 0.912 		& 0.231 		& 0.463 	 \\
			    & (0.116, 0.546)& (0.635, 1.169)& (0.117, 0.331)& (0.321, 0.645)  \\
			    & [68.6\%] 		& [80.0\%] 		& [85.1\%] 		& [66.2\%] \\	
			\midrule
			\multicolumn{5}{l}{ \it Exporting Measure} \\
			$\quad$ intensity	& \checkmark & \checkmark & $\times$ & $\times$ \\
			$\quad$ status	 	& $\times$ & $\times$ & \checkmark & \checkmark \\
			\multicolumn{5}{l}{ \it Pool of Peer Exporters} \\
			$\quad$ same industry	& \checkmark & \checkmark & \checkmark & \checkmark \\
			$\quad$ same region		& \checkmark & $\times$ & \checkmark & $\times$ \\ 
			\midrule
			\multicolumn{5}{p{12.5cm}}{\small \textit{Notes}: Summarized are the results for the LBE and LFE effects obtained using alternative variable specifications. We report the mean point estimates along with 95\% bootstrap percentile confidence intervals (in parentheses) as well as the shares of sample for which the observation-specific estimates are statistically significant at the 5\% level [in brackets]. Model (I) is our main specification, the results for which we reproduce here for convenience.}\\
			\bottomrule[1pt]
		\end{tabular}
	}
\end{table}
% -----------------------------------

As discussed earlier, to identify the spillover effects of exporting beyond exporters, we rule out other potential channels of cross-firm interactions within regions and industries that stem from spatio-industrial proximity of firms to one another. However, we can somewhat relax this assumption by allowing for peer network unobservables that, besides just exporting, may give rise to spillovers in productivity as well (e.g., regional agglomerations). The requirement is that these unobservables be time-invariant, in which case we can control for potential region- and industry-level confounders using peer group fixed effects. We consider group effects across both spatial and industrial dimensions and re-estimate our baseline specification including region and industry fixed effects. The corresponding results are summarized in Table \ref{tab:productivityeffect_robust2}. While predictably the LBE estimates of within-firm effects of exporting hardly change, the mean effect size of external spillovers (LFE) increases notably when we rely solely on within-region or within-region-and-industry variation in the estimation. The increase in the magnitude of spillovers is consistent with significant between-region/industry heterogeneity in (peer) firm exporting which, when omitting group effects, ``dilutes'' the strength of cross-firm effects in the baseline model due to the variation across peer groups.  

Overall, the cross-specification variation in estimates is unsurprising and, in fact, expected because models treat cross-firm interactions a bit differently and/or utilize different variation in data to identify the effects. The qualitative implication however remains unchanged across all alternative specifications: both LBE and LFE are statistically significant and widely prevalent. Also, consistently across these alternatives, we obtain LFE effects that, on average, exceed LBE, implying that the total cross-firm effect of exporting\textemdash \textit{after aggregating across all peers}\textemdash may even out-size the direct learning effect occurring within one firm.\footnote{Analogues of Tables \ref{tab:productivityeffect_robust}--\ref{tab:productivityeffect_robust2} that report \textit{average} indirect LFE effects (per peer) are available in Appendix \ref{sec:appx_addresults}. As with our main specifications, these average indirect effects are much smaller than the direct LBE effect.} If anything, this further underscores the importance of measuring both effects when seeking to quantify the total social effects of exporting (and export-promoting policies, by extension) on domestic industries.

% -----------------------------------
\begin{table}[t]
	\caption{Robustness to Controlling for Peer Group Effects }\label{tab:productivityeffect_robust2}
	\centering\small
	\makebox[\linewidth]{
		\begin{tabular}{l ccc}
			\toprule[1pt]
			& (I) & (II) & (III)  \\
			\midrule
			LBE & 0.363  		& 0.381 		& 0.367 			\\
			& (0.189, 0.555)& (0.202,0.577)& (0.189,0.559) \\
			& [92.7\%] 	& [93.2\%] 	& [92.8\%] 	\\[5pt]
			
			LFE & 0.324  		& 0.711 		& 0.496 		\\
			& (0.116, 0.546)& (0.512,0.958)& (0.236,0.762)  \\
			& [68.6\%] 	& [94.6\%] 		& [83.6\%] 	 \\
			\midrule
			Region FE	& $\times$ & \checkmark & \checkmark  \\
			Industry FE	& $\times$ & $\times$ & \checkmark  \\
			\midrule
			\multicolumn{4}{p{9.2cm}}{\small \textit{Notes}: Summarized are the results for the LBE and LFE effects obtained using alternative fixed-effect specifications. We report the mean point estimates along with 95\% bootstrap percentile confidence intervals (in parentheses) as well as the shares of sample for which the observation-specific estimates are statistically significant at the 5\% level [in brackets]. Model (I) is our main specification, the results for which we reproduce here for convenience.}\\
			\bottomrule[1pt]
		\end{tabular}
	}
\end{table}
% -----------------------------------

\medskip

\textsl{Alternative Two-Step Procedure.}\textemdash Empirical analyses of export-driven spillovers in productivity oftentimes employ a two-step methodology, whereby one first estimates firm productivity via off-the-shelf proxy methods that assume\textemdash and thus impose in the estimation\textemdash an \textit{exogenous} Markov process for productivity and then tests for spillovers in a second step by (linearly) regressing firm productivity estimated in the first step on the export spillover exposure variable(s). For instance, this is the approach taken by \citet{alvarez2008exporting}. As discussed earlier, not only is it internally inconsistent (contradictory) in that a non-zero internal LBE and/or external LFE effect found in the second step \textit{must} violate the key identifying assumption of the first-step estimator, but it is also the case that it is typically implemented using the export spillover exposure measures that average across all firms, including the recipient of spillovers. The latter produces regression coefficients that cannot separate the LBE from the LFE effect, muddying their interpretation.\footnote{For further discussion as well as an illustration for the case of a \citet{alvarez2008exporting}-type second-step regression in \eqref{eq:lopez2008}, see Appendix \ref{sec:appx_b}.}

Notwithstanding these issues, we implement a two-step approach to estimate the LFE effect and compare it with ours. More concretely, we re-estimate our model to obtain firm productivity but, this time, assuming an exogenous Markov process for productivity, whereby $\omega_{it}=h(\omega_{it-1})+\zeta_{it}$ is imposed in place of \eqref{eq:productivitylaw}, and then run the following \citet{alvarez2008exporting}-style second-step linear regression:
\begin{equation}\label{eq:lopez2008}
	E\left[ \widetilde{\omega}_{it} | \cdot \right] = \beta_0 +\beta_{x}X_{it-1}+\beta_{\bar{x}} \overline{X}_{it-1},
\end{equation}
where $\widetilde{\omega}_{it}$ is the first-step estimate of exogenously evolving $\omega_{it}$, and the spillover pool $\overline{X}_{it-1}$ is measured using the ``grand'' average $\sum_{j}p_{iit-1}X_{jt-1}$ that does not exclude $X_{it-1}$. To maximize comparability, we specify the above second-step regression in lags to match the dynamics of \eqref{eq:productivitylaw}.

% -----------------------------------
\begin{figure}[t]
	\centering
	\includegraphics[scale=0.55]{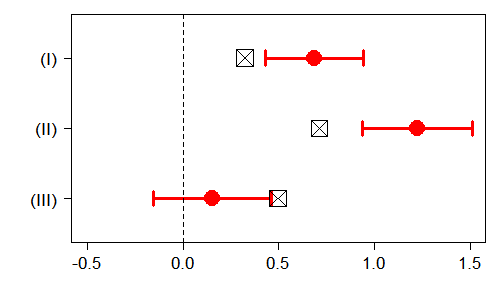}
	\caption{Alternative Two-Step Estimates of the LFE Effect \newline {\small [\textit{Notes}: ``{\Large $\bullet$}'' plots two-step point estimates of the LFE effect from eq.~\eqref{eq:lopez2008}. ``$\boxtimes$'' plots the {mean} point estimates of the LFE effect obtained via our methodology and reported in Table \ref{tab:productivityeffect_robust2}.  Specifications are as follows: (I) -- no fixed effects, (II) -- with region fixed effects, (III) -- with both region and industry fixed effects. Whiskers show 95\% confidence intervals constructed using robust standard errors clustered at the plant level.]}}\label{fig:alt_two_step}
\end{figure}
% -----------------------------------

Figure \ref{fig:alt_two_step} summarizes the estimates of $\beta_{\bar{x}}$ from \eqref{eq:lopez2008} estimated with and without peer group fixed effects. As we show in Appendix \ref{sec:appx_b}, the $\beta_{\bar{x}}$ parameter measures the LFE effect as defined by us in the rest of the paper using the \textit{peer} average measure of $\overline{X}_{it-1}$, viz., $LFE\equiv{\partial E\left[ \omega_{it} | \cdot \right]}\big/{\partial \sum_{j{\ne i}} p_{ijt-1}X_{jt-1}}$.\footnote{The coefficient on the firm's own $X_{it-1}$  is significant and stable across all three specifications (ranging in 0.160--0.168) but we do not focus on it here due to the difficulty interpreting the latter since it does not separately identify the LBE from the LFE.} The three fixed-effects specifications match those reported in Table \ref{tab:productivityeffect_robust2}. 

Estimates from a two-step procedure and our model are non-negligibly different. The specification with both region and industry fixed effects, which drastically reduces the amount of identifying variation in $\overline{X}_{it-1}$ defined at the region-industry level, leads to a noisy estimate of $\beta_{\bar{x}}$ implying an insignificant LFE. Under the other specifications, $\beta_{\bar{x}}$ estimates are all statistically significant but much larger (even exceeding the unit elasticity) than the corresponding mean LFE effects from our model.  Given the lack of internal consistency and the associated misspecification of a productivity process in the first and second steps of the two-step procedure, it is difficult to precisely rationalize why the LFE point estimates differ from ours. Having said that, compared to the two-step procedure, our methodology exhibits a significantly smaller variation in LFE estimates across specifications, lending assurance to their robustness. On the other hand, the dramatic volatility of an \citet{alvarez2008exporting}-style approach is consistent with the simulation evidence reported by \citet{malikovzhao2019} who find that two-step procedures suffer from non-vanishing biases in spillover effect estimates and these biases can be both positive and negative.

% ------------------------------------------------------------------------------------------
% ------------------------------------------------------------------------------------------

\section{Conclusion}\label{sec:conclusion}

%Governments in both the developing and developed countries commonly pursue policies aimed at promoting exports. In addition to boosting aggregate demand, such policies are also routinely justified by arguing that domestic exporters benefit from export-driven productivity improvements. When studying these productivity effects, the existing literature mostly focuses on whether firms improve their performance by engaging in exports \textit{themselves}, a mechanism called ``learning by exporting,'' while largely neglecting a secondary channel whereby domestic firms can also learn from their exporting \textit{peers} via cross-firm spillovers. This indirect learning opportunity, which we refer to as ``learning from exporters,'' is available to both exporters and non-exporters. Omitting this important mechanism may not only provide an incomplete assessment of total productivity benefits of exporting but may also jeopardize the measurement of the more traditional direct learning-by-exporting effects because of the endogeneity-inducing omitted variable bias. 

This paper  extends \citet{de2013detecting} to develop a unified empirical framework for productivity measurement that explicitly accommodates both the direct LBE channel taking place \textit{within} a firm as well as the indirect LFE channel working \textit{between} firms, which enables us to robustly measure and test these two effects in a consistent structural framework. We do so by formalizing the evolution of firm productivity as an export-controlled process, with future productivity potentially affected not only by the firm's own export behavior but also by that of its spatially proximate peers in the industry. This allows a simultaneous, ``internally consistent'' identification of firm productivity and the corresponding effects of exporting. Our identification strategy utilizes a structural link between the parametric production function and the firm's first-order condition for static inputs which helps us circumvent recent non-identification critiques of conventional proxy-based productivity estimators. In addition, owing to a nonparametric treatment of the firm productivity process, our model enables us to accommodate heterogeneity and nonlinearity in productivity effects of exporting across firms.

We apply our semiparametric methodology to a panel of manufacturing plants in Chile in 1995--2007 and find significant evidence of both LBE and LFE. Overall, the LBE effect on productivity is statistically significant for 93\% of firms in our main specification. On average, the size of the LFE effect is commensurate to that of LBE but, at the observation level, the LFE effect is significantly non-zero for 69\% of plants only, thereby suggesting that the indirect productivity-boosting effect of exporting is less prevalent in Chilean manufacturing than the direct learning effect taking place within the firm. We also document that the LFE effect is stronger for firms who also export. This empirical evidence therefore suggests that exporters benefit from the exposure to peer exporters in their local industry more than non-exporters, plausibly because there may be complementarities between internal/direct and external/indirect learning from export experiences. Using simple back-of-envelope calculation, we find that long-run LBE and LFE effects on mean firm productivity are respectively 0.70\% and 0.63\% per percentage point of export intensity. Thus, in the long-run equilibrium a permanent 10 percentage point increase in mean export intensity of all firms in the local industry is roughly estimated to produce a sizable 13.3\% boost to the average firm productivity through both direct LBE and indirect LFE channels.

% ------------------------------------------------------------------------------------------
% ------------------------------------------------------------------------------------------

{\small \setlength{\bibsep}{0pt} \bibliography{main}}

% ------------------------------------------------------------------------------------------
% ------------------------------------------------------------------------------------------

\clearpage
\appendix

\section*{\bf APPENDIX}

\section{Controlling for Self-Selection into Exporting}
\label{sec:appx_selection}

The inclusion of the lagged firm productivity in the productivity process \eqref{eq:productivitylaw} also indirectly allows us to, at least partly, control for self-selection of firms into exporting based on their productivity levels. Consistent with economic theory \citep{melitz2003impact} and vast empirical evidence \citep[e.g.,][]{clerides1998learning,delgado2002firm}, engagement in export markets is more likely among firms with higher productivity levels. However, by including a firm's own lagged productivity in the autoregressive $\omega_{it}$ process, we are able to account (to an extent) for this potential endogenous self-selection when measuring learning effects. That is, both the LBE and LFE effects on future firm productivity in our model are measured after partialling out the contribution of its own productivity, thereby ``fixing'' the past productivity when comparing firms of different export intensities.

Under our model structure, the $X_{it}$ decisions\textemdash while optimized dynamically\textemdash are without delay which allows a contemporaneous correlation between firm productivity $\omega_{it}$ and its export orientation $X_{it}$. This implicitly accommodates the firm's self-selection into exporting based on its productivity. That is, from \eqref{eq:xlaw} and the policy function for $\mathcal{X}_{it}$ implied by the Bellman equation in \eqref{eq:bellman}, we have that $X_{it}(\omega_{it},K_{it},L_{it},X_{it-1})$ whereby the exporting behavior $X_{it}$ contemporaneously and endogenously depends on $\omega_{it}$. 
%our assumption allow for the contemporaneously endogenous dependence of the exporting behavior on productivity:  $X_{it}(\omega_{it})$. 
Substituting for firm productivity using its Markov process, the log production function that we estimate becomes $y_{it}=\alpha_0+\alpha_{K}k_{it}+\alpha_{L}l_{it}+ \beta_Mm_{it}+g(\omega_{it-1},X_{it-1},\overline{X}_{it-1})+ \zeta_{it}+\eta_{it}$. Recognizing that, per our assumptions, $K_{it}(\omega_{it-1},K_{it-1},L_{it-1},X_{it-2})$ and $L_{it}(\omega_{it-1},K_{it-1},L_{it-1},X_{it-2})$, it becomes apparent that our measurement of the effect of $X_{it-1}$ on expected productivity $g(\cdot)$ when estimating the production function is done while controlling for its potential within-firm confounders  $(\omega_{it-1},K_{it}(\omega_{it-1},K_{it-1},L_{it-1},X_{it-2}),L_{it}(\omega_{it-1},K_{it-1},L_{it-1},X_{it-2}))'$ that may impact the determination of $X_{it-1}$.  \citet[][fn.~22]{de2013detecting} makes an analogous argument but in the context of \textit{delayed} exporting decisions.

% ------------------------------------------------------------------------------------------
% ------------------------------------------------------------------------------------------

\section{Using Deflated Revenues to Measure Output}
\label{sec:appx_a}

Beyond just assuming a perfectly competitive output market as commonly done in applied work on productivity, the structural identification outlined in Sections \ref{sec:framework}--\ref{sec:empiricalframework} specifically assumes  that (exogenous) prices are homogeneous across all firms, namely, that $P^Y_{it}=P^Y_t$ for all $i$. The model also implicitly assumes the availability of physical-quantity information about firm outputs in the data. Such information, however, is oftentimes lacking in production datasets, including ours: instead of the output quantity and the corresponding price we only observe the firm's revenue, which is their product. The latter is a trivial point if output prices are indeed common to all firms since, in this case, the output quantity $Y_{it}$ can  be obtained by simply deflating the revenue data using the (observable) output price index $P^Y_t$. However, the assumption that competitive firms are price takers, in theory, does not prevent them from being offered heterogeneous prices. If output prices $P^Y_{it}$ do vary across individual firms and we continue measuring the firm output using deflated revenues in the empirical analysis, the obtained firm productivity will be \textit{revenue}-based, capturing not only the cost/production-side but also the price/demand-side heterogeneity across firms. So long as we maintain the perfect competition assumption in that the firms are price-takers who do not receive ``quantity discounts,'' which renders output prices exogenous to the firm's production decisions, the deviation from the assumed price homogeneity does not invalidate our structural identification arguments except altering the interpretation of a residual TFP as reflecting both idiosyncratic supply and demand heterogeneity at the firm level. In what follows, we show this more formally.

The quantity-based measure of total factor productivity (TFPQ)\textemdash that is, $\omega_{it}$\textemdash appearing in the underlying ``quantity production function'' in \eqref{eq:logprodfun}, $y_{it}=\alpha_{0}+\alpha_{K}k_{it}+\alpha_{L}l_{it}+\alpha_{M}m_{it}+\omega_{it}+\eta_{it}$, captures purely cost/production/supply-side heterogeneity across firms.
When using deflated revenues to measure the firm output, a feasible analogue of this production function estimatable from the data is given by the following ``revenue production function:''
\begin{equation}\label{eq:logrevenueprodfn}
	r_{it}  =\alpha_{0}+\alpha_{K}k_{it}+\alpha_{L}l_{it}+\alpha_{M}m_{it}+\underbrace{\big(\overbrace{\omega_{it}}^{\text{TFPQ}} + \pi_{it}\big)}_{\text{TFPR}} +\eta_{it},
\end{equation}
where $r_{it} = \ln \left( \frac{P_{it}^YY_{it}}{P_t^Y}\right)$ is the deflated revenue, and $\pi_{it}\equiv \ln \left( \frac{P_{it}^Y}{P_t^Y}\right)$ captures heterogeneity in output prices at the firm level as measured by their log-deviation from the common price index. It is easy to see that if $P_{it}^Y=P_{t}^Y$ for all $i$, as in the main case discussed in our paper, $\pi_{it}=0$ and $\text{TFPR}=\text{TFPQ}$ that simplifies \eqref{eq:logrevenueprodfn} to \eqref{eq:logprodfun}. Otherwise, the revenue-based firm productivity TFPR includes both the cost heterogeneity $\omega_{it}$ and the demand heterogeneity $\pi_{it}$. As such, in the context of productivity effects of exporting, the measured TFPR will capture export-driven changes in both the cost and demand factors. Hence, a positive LBE and/or LFE effect can be interpreted as working through either a decline in production cost via the increasing technical productivity or the price-raising increase in demand, or both.

Next, we show that our empirical strategy is generalizable to cases with non-endogenous price heterogeneity across firms, i.e., when $\pi_{it}\ne0$. Without any actual changes in its implementation, our methodology structurally identifies TFPR and LBE/LFE effects thereon if, instead of TFPQ ($\omega_{it}$), we assume that it is the composite TFPR ($\omega_{it}+\pi_{it}$) that is first-order Markov. This revision in assumption is empirically immaterial since the latter continues to formalize the evolution of an ``observationally equivalent'' object, namely, unobservable scalar firm heterogeneity. As discussed earlier, the change is in the interpretation of this heterogeneity only. Thus, we replace \eqref{eq:productivitylaw}  with the following controlled Markov process for TFPR: 
\begin{equation}\label{eq:markovpsi}
	\psi_{it}=h\left(\psi_{it-1},X_{it-1},\overline{X}_{it-1}\right)+\zeta_{it},
\end{equation}
where, for convenience, $\psi_{it}\equiv \omega_{it}+\pi_{it}$. Clearly, \eqref{eq:markovpsi} nests \eqref{eq:productivitylaw} when prices are homogeneous.

Recall that our identification strategy starts with the firm's static optimality condition for materials. The analogue of \eqref{eq:firstorder} for price-taker firms facing heterogeneous output prices $P_{it}^{Y}$ is
\begin{equation}\label{eq:firstorder*}
	\alpha_{M}P_{it}^{Y}A_0K_{it}^{\alpha_{K}}L_{it}^{\alpha_{L}}M_{it}^{\alpha_{M}-1}\exp\left\{ \omega_{it}\right\} \theta=P_{t}^{M}
\end{equation}

Taking the log-ratio of this first-order condition and the underlying ``quantity production function'' of the firm in \eqref{eq:prodfun}, we obtain the same material share equation as in \eqref{eq:share}:
\begin{equation}\label{eq:share*}
	\ln\underbrace{\left(\frac{P_{t}^{M}M_{it}}{P_{it}^{Y}Y_{it}}\right)}_{S_{it}^M}=\ln\left(\alpha_{M}\theta\right)-\eta_{it},
\end{equation}
with the only difference being that the revenue entering the denominator in $S_{it}^M$ now has a firm-specific output price. However, this has \textit{no} bearing on the first stage of our methodology, since $S_{it}^M$ is directly measurable from the data regardless of whether the output price is homogeneous or not. That is, the firm's \textit{revenue} is directly observable under both scenarios, be it equal to $P_{it}^{Y}Y_{it}$ or $P_{t}^{Y}Y_{it}$. As such, material elasticity continues to be identified via $\alpha_{M}=\exp\left\{ E\left[\ln\left(S_{it}^M\right)\right]\right\} \slash E\left[\exp\left\{ E\left[\ln\left(S_{it}^M\right)\right]-\ln\left(S_{it}^M\right)\right\} \right]$ as it was before.

To identify the rest of the production function in the second stage, we now need to work with the ``revenue production function.'' Using the Markov assumption about $\psi_{it}=\omega_{it}+\pi_{it}$ and bringing the already identified material elasticity $\alpha_{M}$ to the left-hand side of the equation, from \eqref{eq:logrevenueprodfn} we obtain
\begin{equation}\label{eq:almostfinal}
	r_{it}^{*}=\alpha_{K}k_{it}+\alpha_{L}l_{it}+\overbrace{g\left(\psi_{it-1},X_{it-1},\overline{X}_{it-1}\right)+\zeta_{it}}^{\psi_{it}}+\eta_{it},
\end{equation}
where $r_{it}^{*}=r_{it}-\alpha_{M}m_{it}$ is fully identified and can be treated as an observable, and $g\left(\cdot\right)\equiv h(\cdot)+\alpha_{0}$ is some function.

To derive the proxy function for the TFPR, we invert the first-order condition in \eqref{eq:firstorder*} after rewriting it slightly (by dividing it through by the common price index $P_t^Y$) as
\begin{equation}
	\alpha_{M}\times\underbrace{\frac{P_{it}^{Y}}{P^Y_t}}_{\exp\{\pi_{it}\}}\times A_0K_{it}^{\alpha_{K}}L_{it}^{\alpha_{L}}M_{it}^{\alpha_{M}-1}\exp\left\{ \omega_{it}\right\} \theta=\frac{P_{t}^{M}}{P_{t}^Y}
\end{equation}
and logging both sides to obtain
\begin{equation}\label{eq:psi*}
	\omega_{it}+\pi_{it} = \underbrace{\ln\left(\frac{P_{t}^{M}}{P_{t}^{Y}}\right)-\ln\left(\alpha_{M}\theta\right)-\left(\alpha_{M}-1\right)m_{it}}_{m_{it}^{*}}-\alpha_{K}k_{it}-\alpha_{L}l_{it}.
\end{equation}
The above material-based proxy for $\psi_{it}$, which includes unobservable cross-firm price heterogeneity  $\pi_{it}$, is now expressed in terms of the observable common price index $P_t^Y$ that we also use when deflating revenues to obtain $r_{it}$. Note that the proxy expression, i.e., the right-hand side of \eqref{eq:psi*} is the same as the one we use in the main body of the paper, but now it controls for a composite heterogeneity $\psi_{it}=\omega_{it}+\pi_{it}$.

The feasible second-stage revenue-based production function is then given by
\begin{equation}\label{eq:final*}
	r_{it}^{*}=\alpha_{K}k_{it}+\alpha_{L}l_{it}+g\left(\left[m_{it-1}^{*}-\alpha_{K}k_{it-1}-\alpha_{L}l_{it-1}\right],X_{it-1},\overline{X}_{it-1}\right)+\zeta_{it}+\eta_{it},
\end{equation}
which identifies $\alpha_{K}$ and $\alpha_{L}$ as well as the LBE/LFE effects on the mean TFPR function $g(\cdot)$ via least squares. The TFPR itself can then be recovered (up to a constant) using the production-function parameters and the shock $\eta_{it}$ from the first stage: $\psi_{it}+\alpha_{0}=r_{it}-\alpha_{K}k_{it}-\alpha_{L}l_{it}-\alpha_{M}m_{it}-\eta_{it}$. 

% ------------------------------------------------------------------------------------------
% ------------------------------------------------------------------------------------------

\section{Bootstrap}
\label{sec:appx_boot}

For statistical inference, we employ accelerated biased-corrected percentile bootstrap confidence intervals proposed by \citet{efron1987better}, which can correct for finite-sample biases and control for higher moments (skewness) of the sampling distribution. Due to the panel structure of data, we employ wild residual block bootstrap, which can preserve the within-firm correlation in the data, to approximate the sampling distribution of the estimator. In addition, we bootstrap both stages jointly because the estimation in the second stage is based on the first-stage estimator.

Having obtained the bootstrap
estimates of all parameters $\big\{ \left(\widehat{\alpha}_{K}^{b},\widehat{\alpha}_{L}^{b},\widehat{\alpha}_{M}^{b},\widehat{\boldsymbol{\gamma}}^{b}\right)';\ b=1,2,...B\big\}$, we then use them to construct the bootstrap replications of our main estimands of interest:
$\big\{\widehat{LBE}_{it}^b\big\}$ and $\big\{\widehat{LFE}_{it}^b\big\}$, which we then use to  construct accelerated biased-corrected percentile confidence intervals for each observation-specific $LBE_{it}$ and $LFE_{it}$. We use $B=500$ bootstrap replications. 

For simplicity, let $\widehat{Z}$ represent an estimate of some functional of interest $Z$ and $\{\widehat{Z}^{b}\}$ be the set of its bootstrap estimates. Then, a two-sided $(1-\alpha)100\%$ confidence intervals for
$Z$ is the $\left[\alpha_{1}\times100\right]$th and $\left[\alpha_{2}\times100\right]$th
percentiles of the empirical distribution of $\{\widehat{Z}^{b}\}$, where
\begin{equation*}
	\alpha_{1}=\Phi\Bigg(\widehat{\lowercase{q}}_{0}+\frac{\widehat{\lowercase{q}}_{0}+\lowercase{q}_{\alpha/2}}{1-\widehat{c}\left(\widehat{\lowercase{q}}_{0}+\lowercase{q}_{\alpha/2}\right)}\Bigg),\quad 
	\alpha_{2}=\Phi\Bigg(\widehat{\lowercase{q}}_{0}+\frac{\widehat{\lowercase{q}}_{0}+\lowercase{q}_{\left(1-\alpha/2\right)}}{1-\widehat{c}\left(\widehat{\lowercase{q}}_{0}+\lowercase{q}_{\left(1-\alpha/2\right)}\right)}\Bigg),
\end{equation*}
and $\Phi$ denotes the standard normal cdf; $\lowercase{q}_{\alpha/2}=\Phi^{-1}(\alpha/2)$; $\widehat{\lowercase{q}}_{0}$ denotes a bias-correction factor defined as $\widehat{\lowercase{q}}_{0}=\Phi^{-1}\left({\#\left\{ \widehat{Z}^{b}<\widehat{Z}\right\} } \big/{B}\right)$; 
$\widehat{c}$ denotes an acceleration parameter which, following the literature \citep[][]{shao2012jackknife}, we estimate via jackknife: $\widehat{c}={\sum_{j=1}^{J}\left(\sum_{s=1}^{J}\widehat{Z}^{s}-\widehat{Z}^{j}\right)^3}\big/ \big\{6\big[\sum_{j=1}^{J}\left(\sum_{s=1}^{J}\widehat{Z}^{s}-\widehat{Z}^{j}\right)^{2}\big]^{3/2}\big\}$,
where $\big\{\widehat{Z}^{j}\big\}$ are the jackknife estimates of $Z$.\footnote{To account for the panel structure of data and to manage computational time, we use a delete-$20T$ jackknife, i.e., we leave 20 cross-sections out to obtain jackknife estimates. }

% ------------------------------------------------------------------------------------------
% ------------------------------------------------------------------------------------------

\clearpage
\section{Data Summary Statistics}
\label{sec:appx_data}

\setcounter{table}{0}
\renewcommand\thetable{C.\arabic{table}}

% -----------------------------------
\begin{table}[h]
	\caption{Data Summary Statistics, 1995--2007}\label{tab:data}
	\centering\small
	\begin{threeparttable}
		\begin{tabular}{lrrrr}
			\toprule[1pt]
			Variable & Mean & 5th Perc. & Median & 95th Perc.\\
			\midrule
			Output ($Y$) & 276.46  & 8.37  & 54.75  & 1,348.35 \\
			Capital ($K$) & 102.24  & 0.48  & 12.15  & 513.97 \\
			Labor ($L$) & 49.74  & 6.00  & 22.00  & 196.00 \\
			Materials ($M$) & 125.94  & 2.85  & 24.09  & 626.56 \\
			Export Intensity ($X$) & 0.056  & 0.000  & 0.000  & 0.499 \\
			Exposure to Exporter ($\overline{X}$) & 0.060  & 0.000  & 0.036  & 0.217 \\
			%Exporter Status & 0.206  &  &  & \\
			\midrule
			\multicolumn{5}{p{10.5cm}}{\small \textit{Notes}: $Y$, $K$, $M$ are measured in hundred thousands of real pesos. $L$ is measured in the number of people. $X$ and $\overline{X}$ are unit-free proportions.}\\
			\bottomrule[1pt]
		\end{tabular}
		
	\end{threeparttable}
\end{table}
% -----------------------------------

% ------------------------------------------------------------------------------------------
% ------------------------------------------------------------------------------------------

\section{Additional Results}
\label{sec:appx_addresults}

\setcounter{table}{0}
\renewcommand\thetable{D.\arabic{table}}

% -----------------------------------
\begin{table}[h]
	\caption{Analogue of Table \ref{tab:productivityeffect_robust}}
	\centering\small
	\makebox[\linewidth]{
		\begin{tabular}{l cccc}
			\toprule[1pt]
			& (I) & (II) & (III) & (IV) \\
			\midrule
			(direct) LBE & 0.363  		&  0.379 		& 0.147 		& 0.151 		 \\
				& (0.189, 0.555)& (0.207, 0.575)& (0.117, 0.176)& (0.121, 0.179) \\[5pt]
			
			(indirect) LFE$\quad$  	& 0.013  		& 0.004 		& 0.009 		& 0.003 	 \\
			from one peer & (0.005, 0.022) & (0.003, 0.005) & (0.004, 0.013) & (0.002, 0.003)  \\	
			\midrule
			\multicolumn{5}{p{12.3cm}}{\small \textit{Notes}: This table reproduces the results reported in Table \ref{tab:productivityeffect_robust} except that indirect LFE effects are computed \textit{per peer}. Presented are the mean point estimates along with 95\% bootstrap confidence intervals (in parentheses). Model (I) is our main specification.}\\
			\bottomrule[1pt]
		\end{tabular}
	}
\end{table}
% -----------------------------------

% -----------------------------------
\begin{table}[h]
	\caption{Analogue of Table \ref{tab:productivityeffect_robust2}}
	\centering\small
	\makebox[\linewidth]{
		\begin{tabular}{l ccc}
			\toprule[1pt]
			& (I) & (II) & (III)  \\
			\midrule
			(direct) LBE & 0.363  		& 0.381 		& 0.367 			\\
			& (0.189, 0.555)& (0.202,0.577)& (0.189,0.559) \\[5pt]
			
			(indirect) LFE$\quad$ & 0.013  		& 0.030 		& 0.020 		\\
			from one peer & (0.005, 0.022) & (0.021, 0.040) & (0.010, 0.031) \\

			\midrule
			\multicolumn{4}{p{10cm}}{\small \textit{Notes}: This table reproduces the results reported in Table \ref{tab:productivityeffect_robust2} except that indirect LFE effects are computed \textit{per peer}. Presented are the mean point estimates along with 95\% bootstrap confidence intervals (in parentheses). Model (I) is our main specification.}\\
			\bottomrule[1pt]
		\end{tabular}
	}
\end{table}
% -----------------------------------

% ------------------------------------------------------------------------------------------
% ------------------------------------------------------------------------------------------

\clearpage
\section{Using ``Grand'' Averages to Measure the Spillover Exposure}
\label{sec:appx_b}

As discussed in Section \ref{sec:framework} and specifically shown in \eqref{eq:lbelfeident}, by measuring the $i$th firm's exposure to exporters using the average \textit{peer} export orientation $\sum_{j{\ne i}} p_{ijt}X_{jt}$ which excludes exports of the firm $i$ ``receiving'' spillovers we are able to separately identify LBE from LFE. This would not be the case if the export spillover exposure was measured using the ``grand'' average export orientation $\sum_{j} p_{ijt}X_{jt}$ that aggregates over all units in the peer group, including the firm $i$. We illustrate this point for a second-step regression of the \citet{alvarez2008exporting}-style approach to testing for export spillovers given in \eqref{eq:lopez2008}:
\begin{equation*}
	E\left[ \widetilde{\omega}_{it} | \cdot \right] = \beta_0 +\beta_{x}X_{it-1}+\beta_{\bar{x}} \overline{X}_{it-1}. 
\end{equation*}
In what follows, we abstract away from the identification problems associated with measuring $\widetilde{\omega}_{it}$ in the first step without accounting for its relation with $X_{it-1}$ and $\overline{X}_{it-1}$.

Let us first note that, had the \textit{peer} average been used to measure $\overline{X}_{it}$, the regression coefficients in \eqref{eq:lopez2008} would have identified both LBE and LFE effects. Namely, from
\begin{align}
	E\left[ \widetilde{\omega}_{it} | \cdot \right]  &= \beta_0 + \beta_{x}X_{it-1}+\beta_{\bar{x}} \Big( \sum_{j{\ne i}} p_{ijt-1}X_{jt-1}\Big) 
\end{align}
we have that
\begin{align*}
	LBE&\equiv\frac{\partial E\left[ \widetilde{\omega}_{it} | \cdot \right]}{\partial X_{it-1}} = \beta_{x} \qquad \text{and } \qquad LFE\equiv\frac{\partial E\left[ \widetilde{\omega}_{it} | \cdot \right]}{\partial \sum_{j{\ne i}} p_{ijt-1}X_{jt-1}} = \beta_{\bar{x}}.
\end{align*}

However, because $\overline{X}_{it}$ is measured using the ``grand'' average export intensity, the regression coefficients in \eqref{eq:lopez2008} do not separately identify LBE from LFE, viz., from 
\begin{align}
	E\left[ \widetilde{\omega}_{it} | \cdot \right] &= \beta_0 + \beta_{x}X_{it-1}+\beta_{\bar{x}} \Big( \sum_{j} p_{ijt-1}X_{jt-1}\Big) \notag \\ 
	&= \beta_0 + \beta_{x}X_{it-1}+\beta_{\bar{x}} \Big( p_{iit-1}X_{it-1} + \sum_{j{\ne i}} p_{ijt-1}X_{jt-1}\Big)  \label{eq:lopez2008_grand}
\end{align}
it immediately follows that
\begin{align*}
	LBE&\equiv\frac{\partial E\left[ \widetilde{\omega}_{it} | \cdot \right]}{\partial X_{it-1}} = \beta_{x} + \beta_{\bar{x}} p_{iit-1} \ne \beta_{x}. 
\end{align*}

The above shows that the coefficient on the firm's own exporting $X_{it-1}$ in \eqref{eq:lopez2008_grand} does \textit{not} identify the LBE effect, albeit that is how it is typically interpreted in such two-step analyses. This cross-contamination of the effects vanishes only asymptotically as the peer group size diverges, forcing $p_{iit-1}\to 0$. This is a nonissue in our case because, when using the \textit{peer} average $\sum_{j{\ne i}} p_{ijt-1}X_{jt-1}$, we \textit{a priori} set $p_{iit-1}=0$. 

Interestingly, the second-step regression in \eqref{eq:lopez2008_grand} does identify the LFE effect, but as we define it in our paper:
\begin{align*}
	LFE&\equiv\frac{\partial E\left[ \widetilde{\omega}_{it} | \cdot \right]}{\partial \sum_{j{\ne i}} p_{ijt-1}X_{jt-1}} = \beta_{\bar{x}}, 
\end{align*}
i.e., as the effect of a marginal change in the \textit{peers'} average exporting. This is different from the typical conceptualization of an external cross-firm ``spillover effect'' in two-step analyses that use ``grand'' average $\overline{X}$, viz., $SPILL\equiv{\partial E\left[ \widetilde{\omega}_{it} | \cdot \right]}\big/{\partial \sum_{j} p_{ijt-1}X_{jt-1}}$. It is easy to see that the regression coefficient on $\overline{X}_{it-1}$ in \eqref{eq:lopez2008_grand} does \textit{not} identify the $SPILL$ effect either. Specifically, from 
\begin{align}
	E\left[ \widetilde{\omega}_{it} | \cdot \right] &= \beta_0 + \beta_{x}X_{it-1}+\beta_{\bar{x}} \Big( \sum_{j} p_{ijt-1}X_{jt-1}\Big) \notag \\
	&= \beta_0 + \frac{\beta_{x}}{p_{iit-1}}\Big( \sum_{j} p_{ijt-1}X_{jt-1} - \sum_{j{\ne i}} p_{ijt-1}X_{jt-1}\Big) +\beta_{\bar{x}} \Big( \sum_{j} p_{ijt-1}X_{jt-1}\Big) 
\end{align}
we get
\begin{align*}
	SPILL\equiv\frac{\partial E\left[ \widetilde{\omega}_{it} | \cdot \right]}{\partial \sum_{j} p_{ijt-1}X_{jt-1}} = \frac{\beta_{x}}{p_{iit-1}} + \beta_{\bar{x}} \ne \beta_{\bar{x}} .
\end{align*}

The above illustrates that not only is the $SPILL$ effect not identified by the regression parameter $\beta_{\bar{x}}$ but it is also not well-defined to begin with, because it explodes as the peer group size grows.
As such, both regression coefficients $\beta_{x}$ and $\beta_{\bar{x}}$ from the model estimated using the ``grand'' average $\overline{X}$ lack a meaningful interpretation within the model's own conceptual paradigm.

\end{document}